\documentclass[12pt]{article}
\usepackage{amssymb}
\usepackage{amsmath}
\usepackage{stmaryrd}
\usepackage{tabularx}
\usepackage{hyperref}
 \usepackage{ulem}
\def\theequation {\thesection.\arabic{equation}}
\makeatletter\@addtoreset {equation}{section}\makeatother

\def\ardlabel#1{\let\@currentlabel=\theequation\label{#1}}

\newtheorem{theorem}{Theorem}[section]

\newtheorem{proposition}[theorem]{Proposition}

\newtheorem{remark}[theorem]{Remark}

\usepackage[dvips]{epsfig}
\usepackage{graphicx}
\usepackage[utf8]{inputenc}
\usepackage[english]{babel}
\usepackage[dvipsnames]{xcolor}

\definecolor{mypink3}{cmyk}{0, 0.7808, 0.4429, 0.1412}
\definecolor{mygray}{gray}{0.6}

\begin{document}

\title{\bf Water wave problem with inclined walls}

\author{ 
P. Panayotaros$^*$, R.M. Vargas-Maga\~na$^{**}$, \\
{$^*$  \small Depto. Matem\'{a}ticas y Mec\'{a}nica,
I.I.M.A.S.-U.N.A.M., } \\
{\small Apdo. Postal 20--726, 01000 Ciudad de M\'{e}xico, M\'{e}xico,} \\ 
{ \small $^{**}$ School of Mathematics, University of Edinburgh}, \\
{\small Edinburgh, Scotland, EH9 3FD, UK}
}

\date{\today}

\maketitle

\begin{abstract}

We study free surface water waves in a 2-D symmetric triangular channel
with sides that have a $45^o$ slope. We develop models
for small amplitude nonlinear waves, extending earlier studies that 
have considered the linearized problem. We see that a combination of heuristic
small amplitude expansions lead to a relatively simple system that we then use to 
study interactions of low frequency modes. 
The formalism relies on an explicit construction of the normal modes of the linear problem
and a new way to represent the free surface. We argue that the construction 
can be applied to more general geometries. 
We also examine the structure and some dynamical features 
of spectral truncations 
for the lowest even modes.  


\end{abstract}

\line(1,0){458}


\section{Introduction}
\label{intro}

We propose a simplified nonlinear model for  
free surface potential flow in 2-D domains with slopping lateral rigid wall boundaries. 
i.e. ``beach geometries''. The paper is partially motivated by 
classical studies of the linearized system, 
especially works on special geometries where the linear normal modes can be found in explicit or semi-explicit ways,  
such as isosceles triangles with sides inclined at $45^o$ and $60^o$ to the vertical, see 
\cite{K79,KH80, G87,M93,L32,P80,G94,G95,VMMP19}, and semicircular channels \cite{EL93}. The theory 
extends to 3-D channels with constant cross sections given by the 
above shapes \cite{M93,G94,G95}. 
The suitability of the potential flow model for problems in such geometries may be questioned on physical grounds,
and there are alternative models currently in use that incorporate other effects, see e.g. \cite{HN81,LL89,KMT91,BB08,TFPP13}. 
On the other hand the potential flow theory has made considerable advances into the 
related problem of waves in variable depth \cite{CGNS05,L13,AP17,AN18}, and the present work considers some of   
the geometrical problems involved in approximating the potential flow system in inclined wall geometries. Our main goal is to obtain a simple weakly nonlinear model that can be also generalized to other geometries. Our strategy is to work almost 
exclusively with the $45^o$ isosceles triangle domain \cite{K79,KH80,G87,L32}. We see that the derivation of a weakly nonlinear system points to possible extensions to more general domains, although workable models for general domains involve nontrivial additional ingredients that are left for future work. 

The main feature of our approach is a new 
description of the free surface as the image of the  
flat surface under the flow of a 
vector field that is the gradient of a harmonic function with  
vanishing normal derivative at the rigid walls.   
The dynamical variables of the problem are then   
the usual velocity potential, and this second harmonic 
function used to compute the free surface. 
We show that this description 
of the system, together with the Euler equations for potential flow, can be used 
to derive a simpler approximate model equation for weakly nonlinear waves. 
The derivation also relies on an explicit small amplitude approximation  
of the free surface and uses expansions in the normal modes of the linear problem.  
The goal of the paper is to show how this formulation can be used to compute  
dynamical quantities of special interest such as wave run-up on a beach.

The proposed representation of the free surface 
is motivated by limitations of the simpler graph description in domains 
with inclined domains and is 
described in section 2,  
where we also discuss its main properties, advantages, and disadvantages.  
Our definition implies that the free surface   
satisfies automatically some desirable geometrical properties, 
such as area conservation, and non self-intersection. 
Also, the end points of the free surface are always on the rigid wall.
Such properties should follow from the dynamics of the fluid, but  
are here imposed by the representation of the free surface. 
This allows for more drastic approximations 
of the dynamical equations.    
A drawback of the proposed representation is that  
in general there is no explicit exact expression of the free surface   
in terms of the auxiliary potential. Despite this fact,     
the existence and properties of the curve representing free surface 
can be deduced using basic facts from the 
theory of ordinary differential equations, see e.g. \cite{P91}.     
Most importantly, 
the free surface can be expressed explicitly in terms of the auxiliary 
potential using approximate formulas that can be 
interpreted as small amplitude (surface deformation) 
expansions.  
We show that 
the geometrical properties of the exact curves are also satisfied      
by such approximations, 
either exactly or approximately, 
with errors that we can quantify. 
We also point out a simple approximation, related to Euler's formula for numerical integration,    
that is especially useful for domains with constant slope.

We then show that the proposed description of the free surface,   
its small deformation approximation by Euler's formula, and the equations for free surface potential flow 
can be combined to derive an approximate equation for weakly nonlinear waves. 
The first step of this derivation leads to  
the linear evolution equations appearing in the classical results on normal modes cited above.
The results on the linear problem are then used to derive a lowest order (quadratic) nonlinear model 
for the interactions 
of the normal modes of the linearized problem. 
The construction
of the spectral equation uses the results of \cite{K79,KH80,G87,L32} on linear normal modes of the $45^o$ triangular domain. 
The normal modes form a set of functions that are harmonic in the domain 
that can be occupied by the fluid and also satisfy the rigid wall boundary conditions. Similar functions are known in  
simpler geometries, e.g. the constant depth channel, and 
are also used in the computation of the Dirichlet-Neumann operator and the derivation 
of approximate models \cite{CS93}, see also \cite{WV15} for comparison of different methods.    
The paper shows that 
the tools used to analyze 
the Dirichlet-Neumann operator
can be used to describe the free surface in domains with inclined walls. 

We also argue that  
the derivation of the nonlinear model equation can be meaningful for general domains with slopping beaches. 
This equation is potentially interesting for its structure, but in the absence of explicit expressions for the normal modes 
its numerical study will require alternative discretizations. 
The model also involves the Dirichlet-Neumann operator for the flat surface domain
and related regularity questions that require some care \cite{MW16}. These questions 
are more tractable in the special geometry we consider as we work with linear combinations of functions that are harmonic
in domains that include the fluid domain. We note that quadratic water wave models with full dispersion (e.g. ``Whitham-Boussinesq'' systems) \cite{AMP13,MKD15,H17,HT18} can capture several effects of more involved models see e.g. \cite{VP16,C18,VMS21}, especially in the long wave regime.

We apply the quadratic model to study mode interactions in the $45^o$ triangular domain. We examine 
the evolution on the 
invariant subspace of even modes and present numerical experiments of nonlinear mode interactions, and 
the computation of 
the corresponding wave amplitudes and run-up. 
We focus on near-monochromatic initial conditions, and we gradually increase the amplitude to see 
nonlinear mode interactions leading to amplitude and phase modulation of the dominant mode. 
Some solutions with two modes of comparable amplitude are also presented. 
In all cases larger initial conditions lead to rapid growth of the amplitude and we focus on results suggesting 
long time existence for small amplitudes. Nonlinear mode interactions lead to marked effects on wave run-up at the beach.   
The long time dynamics of the truncated mode interactions can be examined in more detail in future work and we comment 
on some relevant issues in the discussion section. 
The construction of the mode interaction systems involves the computation of mode interaction coefficients, 
details are presented in a Supplement.

In section 2 we present the general formalism, the derivation of an approximate model for the $45^o$ domain
and discuss generalizations. In section 3 we present some computations with the 
quadratic spectral equations on the invariant space of even modes.  
In section 4 we discuss some limitations of the present study and possible further work.  
Computations of mode interaction coefficients are detailed in a Supplement.

\section{Free surface formulation and weakly nonlinear model}
  
In this section we introduce the description of the free surface by 
a potential in its exact and approximate form. 
We use the equations of free surface potential flow to 
show how this description leads to  
the well known problem of linear modes. We further use
the approximate description of the free surface and the analysis of the linearized problem 
to  derive an approximate  weakly nonlinear model and write it in spectral form. 
We discuss generalizations and further properties and limitations 
of the proposed description of the free surface.

We consider free surface potential flow in two dimensional 
domain with rigid walls at
\begin{equation}
  \partial {\cal W} = \{ [x,y] \in \mathbb{R}^2: y = |x|\}.
  \label{forty-five-wall}
\end{equation}
The fluid will occupy a time-dependent domain $D_t$ contained in the set 
${\cal W} = \{ [x,y] \in \mathbb{R}^2: y \geq |x|\}$.
The fluid surface at rest is at $y = B > 0$.

The free surface
at time $t$ can be parametrized by a curve $[X(s,t), Y(s,t)]$,
$s \in [-h,h]$. Such a curve is assumed to not self-intersect, to 
intersect the boundary $\partial {\cal W}$ only at $s = \pm h$,  
and to satisfy boundary conditions
$- X(-h,t) = Y(-h,t)$ and $ X(h,t) = Y(h,t)$. By  
  conservation of fluid volume it should also satisfy
  \begin{equation}
    \label{volume-conservation}
 \frac{1}{2} \int_{-h}^{h}(Y(s,t)\;dX - X(s,t)\;dY)  
=  B^2.
\end{equation}
We expect that the evolution preserves these conditions.  

\subsection{\it Description of the free surface}

We propose a more restricted but less explicit way to
parametrize the free surface as 
\begin{equation}
[X(s,t), Y(s,t)] = {\cal G}^\epsilon(s,t) \in \mathbb{R}^2,
\quad s \in  [-h,h],
\label{vector-param-surf}
\end{equation}
$\epsilon > 0$ a constant, 
where $  {\cal G}^\epsilon(s,t)$ 
is the image of the set 
$\{[x,y] \in \mathbb{R}^2: x \in [-h,h], y = B \} $, 
the free surface at rest, 
under the  ``time$-\epsilon$ flow'' (see next paragraph for this terminology)
of a vector field of the form
$ \nabla V(\cdot,t)$ in $ {\cal W}$ 
with $V(\cdot,t)$ a function that is harmonic in  ${\cal W}$ 
and satisfies the Neumann condition 
$ {\hat n} \cdot \nabla V(\cdot,t) = 0$ at
$\partial {\cal W}$,  $\hat n$ the normal vector at $\partial {\cal W}$, 
for all times $t$. 

The above definition uses some basic facts and terminology from the theory of 
ordinary differential equations, and we provide some additional details.
Specifically, for each time $t$, fix $t$ and  
consider the ordinary differential equation 
\begin{equation}
\label{surface-flow}
 \frac{d r}{d\tau} = \nabla V(r,t), \quad r \in {\cal W}.  
 \end{equation}
The variable $\tau$ is not the physical time but rather the variable used to parametrize  
the integral curves of the vector field  $\nabla V(\cdot,t)$ at each fixed time $t$, also  
$r = [x,y]$ are coordinates of points in $\cal W$. 
To construct ${\cal G}^\epsilon(\cdot,t)$
we integrate \eqref{surface-flow} in $\tau$,  then ${\cal G}^\epsilon(s,t)$ is the unique solution $r(\tau,s;t)$ of  \eqref{surface-flow}
at $\tau = \epsilon$ that satisfies 
$r(\tau) = [s,B]$ at $\tau = 0$. 
(The ``time$-\epsilon$ map'' of \eqref{surface-flow} is the map from the 
$r(0) = [s,B]$ to the $r(\epsilon,s;t)$, with $s \in [-h,h]$.)  
The existence and uniqueness of such 
$r(\tau,s;t)$ for all $s \in [-h,h]$ at any given $t$ follows from the theory of existence and uniqueness for ordinary differential equations 
under certain (continuity, differentiabilty, etc.) assumptions on the vector field $\nabla V(\cdot,t)$, see e.g. \cite{P91}, ch. 2, 
and [L07], ch. 5. We assume that these conditions 
are satisfied for all $t$. This can be checked   
once we have $V(\cdot,t)$, see Remark \ref{local-existence} for some details. Figure 1 shows a schematic diagram of map  ${\cal G}^\epsilon(s,t)$ at time $t$ with $s \in [-h,h].$

Thus the dynamical variable behind this representation 
the free surface is $V(\cdot,t)$, 
i.e. $V(\cdot,t)$ determines 
${\cal G}^\epsilon(s,t) $ uniquely, for all $s \in [-h,h]$ and $t$.
The dynamical equation for $V$ are derived in the next subsections. 
Note that the definition of ${\cal G}^\epsilon(\cdot,t)$ is abstract, in that we do not have 
an explicit expression for  ${\cal G}^\epsilon(\cdot,t) $ in terms of $V(\cdot,t)$. 
We emphasize however that for $\epsilon$ small, 
the solution of \eqref{surface-flow} can be approximated by explicit formulas, i.e. expressions that are given in terms of $V$. 
(Such formulas are used 
in numerical integrators, see e.g. \cite{L07}.) 
One such approximation, Euler's formula, is defined in subsection 2.4 below, and 
used in the derivation of approximate equations for $V$ in subsections 2.5-2.7. 
 
The representation of the free surface by ${\cal G}^\epsilon$ implies that the curve representing the free surface satisfies 
a number of desirable properties. 
First, \eqref{vector-param-surf}, \eqref{surface-flow} imply 
area conservation, and \eqref{volume-conservation}, at all times $t$. 
This follows by $V$ harmonic in ${\cal W}$,  
then the divergence of $ \nabla V$ vanishes, and we have volume conservation
by a standard argument, see also \cite{P91}, ch. 2. 
Furthermore, ${\cal G}^\epsilon$ maps the points $s = \pm h$ to points on $\partial {\cal W}$ at all times $t$, 
i.e. these two points are always on the wall. This follows from the fact that $\nabla V$ is tangent to $\partial  {\cal W}$, 
at all times. Then the integral curves of \eqref{surface-flow} passing from any point on $\partial {\cal W}$ stay on $\partial {\cal W}$.    
Note that ${\cal G}^\epsilon(\pm h,t)$ are the points where the free surface intersects the rigid wall.  
We also check that these are the only points where the free surface ${\cal G}^\epsilon(s,t)$, $s \in [-h,h]$, can intersect the 
rigid wall: any point of ${\cal G}^\epsilon(s_*,t)$, $s \in (-h,h)$, in $\partial {\cal W}$, must satisfy 
${\cal G}^0(s_*,t)  = s_* \in \partial {\cal W}$ by the invariance of $\partial {\cal W}$ under the flow of $\nabla V(\cdot,t)$.  
That would imply $s_* \in \{-h,h\}$, a contradiction. 
Also, by 
the theory of existence and uniqueness of the initial value problem for  \eqref{surface-flow}, 
the function $[s,B] \mapsto {\cal G}^\epsilon(s,t) $, $s \in [-h,h]$, defining the free surface 
is invertible \cite{P91}. This fact 
implies that the free surface can not self-intersect, i.e. there are no $-h< s_1 < s_2< h$ and $t$ for which 
$ {\cal G}^\epsilon(s_1,t) = {\cal G}^\epsilon(s_2,t)$.


\begin{figure}
    \centering
     \includegraphics[angle=270,scale=0.34]{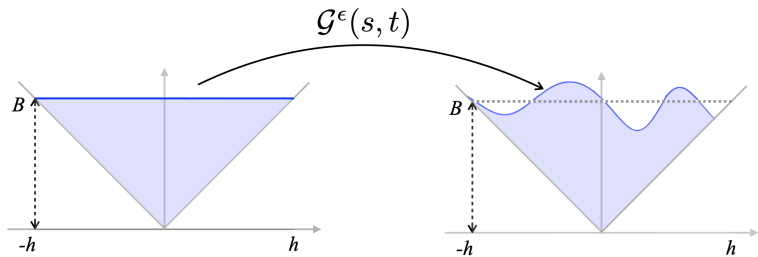}
    \caption{The free surface at time $t$ is represented 
    by a map ${\cal G}^\epsilon(s,t)$, $s \in [-h,h]$, of the blue line on the left to the blue curve on the right. 
    ${\cal G}^\epsilon$ is constructed by fixing $t$ and integrating \eqref{surface-flow} in $\tau$ from $\tau = 0 $ to 
    $\tau = \epsilon$ with initial conditions the points on the blue curve on the left (the flat surface). The map ${\cal G}^\epsilon$
    can be approximated by explicit expressions.}   
    \label{fig:p001}
\end{figure}


The proposed description of the free surface is motivated by the difficulty in using the 
usual simpler definition of the free surface as a graph, i.e. as the set of  points with 
cartesian coordinates $[x, \eta(x,t)]$. The problem in a fluid domain with inclined walls is 
that the domain of the horizontal coordinate $x$ will generally vary in time. 
Many works, including the cited literature on linear normal modes, do not address this 
issue, but still obtain meaningful results, see e.g. \cite{L32}. 
However the point where the free surface 
intersects the rigid wall is defined in an ad-hoc way, e.g. the normal mode equation is defined for 
$x$ is in a fixed domain, the free surface at rest,  
and the domain of $\eta$ is extended to a larger set. This allows us to 
determine the intersection of the graph of this extended $\eta$ and the rigid wall.
More importantly, 
the absence of a clear definition of the free surface in the classical linear literature  
is an obstacle to deriving weakly nonlinear equations.

The use of systems of coordinates where the rigid wall is the vertical direction is possible, 
but this approach requires different systems for each domain. Also the height of fluid at rest is not constant.
The work of \cite{C17} discusses more general parametrizations in two dimensions
that can be also used for domains with inclined walls, especially 
parametrization by arc length. These parametrizations  
do not satisfy automatically the 
geometrical properties of ${\cal G^\epsilon}$ discussed above, e.g. area conservation and permanence of the end points 
on the wall relies on the dynamics. 
The proposed method can be also applied to three dimensional domains, e.g. using the classical normal mode results  
for some three dimensional channels with constant cross section \cite{G94,G95}.  
An alternative more general description of the free surface that uses level sets 
has been implemented numerically and used extensively in coastal engineering, 
using Navier-Stokes (and versions with turbulent Reynolds stress terms) \cite{HN81,LL89,KMT91,BB08,TFPP13}. 
These equations can not be reduced to equations for surface quantities. Independently of this fact,   
the level set approach is more complicated as it adds a partial differential equation for the evolution of 
the fluid domain,      
and may not easily lead to the kind of more quantitative analysis that seems possible with our model. 
Further comments on the advantages and disadvantages of the proposed description of the free surface
at the end of this section, and in the discussion section.

We now discuss the dynamics of the free surface system. We show that we can combine 
the description of the free surface above,  an approximation of ${\cal G}^\epsilon$ defined below, and 
the equations of motion for free surface potential flow 
to obtain approximate equations for the evolution of the 
free surface and the velocity inside the fluid. 


\subsection{ \it Dynamical variables of the system}

A preliminary step is to identify the dynamical variables of the problem. 
These will be the potential $V(\cdot,t)$ used to define the free surface via \eqref{surface-flow}, 
and the hydrodynamic potential $\phi(\cdot,t)$. 
Both quantities will be also represented by an expansion in a set of harmonic functions $\Phi_n$
that will be given explicitly below. They appear in the analysis of the linearized system below.

First,  
the set of possible harmonic functions $V(\cdot,t) $ 
will be written as
\begin{equation}
  V(\cdot,t) = \sum_{n=1}^{\infty}b_n(t) \Phi_n(\cdot),
  \label{psi-expansion}
\end{equation}
where the functions $\Phi_n $, to be 
specified below, are harmonic in ${\cal W}$
and satisfy 
$ \partial_{\hat n}\Phi_n = 0$ at $\partial {\cal W}$.

In free surface potential flow, the Eulerian velocity $u$ in the
fluid domain will be $u = \nabla \phi$, with
$\phi$ harmonic in $\cal W$ with vanishing normal derivative  
at $\partial {\cal W}$.
We will expand  $\phi(\cdot,t) $ as 
\begin{equation}
  \phi(\cdot,t) = \sum_{n=1}^{\infty}a_n(t) \Phi_n(\cdot).
  \label{phi-expansion}
\end{equation}
By the properties of the $\Phi_n$,
$ \phi(\cdot,t)$ is harmonic in ${\cal W}$
and satisfies 
$ \partial_{\hat n}\phi = 0$ at $\partial {\cal W}$.

\begin{remark}
\label{local-existence}
(i) One of the ideas implicit on the above expansions is that the descriptions 
of $V$ (hence the free surface), and $\phi$ involve similar tools. 
(ii) We see below that the functions $\Phi_n$ used in the $45^0$ slope domain are analytic on the whole plain. 
The properties of $V$
needed to guarantee the existence of ${\cal G}^\epsilon$,  
e.g. the size of the derivative of $\nabla V$ in a neighborhood 
of the triangle bounded by $\cal W$ and points with height $y < B$, see e.g. \cite{L07}, ch. 5, 
are controlled by the size of 
the $b_n$.  
The nontrivial information is the evolution of the $b_n$. Truncations to finite sums used in the next section 
again make these technicalities much more tractable as long at the $b_n$ remain bounded.    
\end{remark}

To derive equations for the evolution of $V$, $\phi$ we use Euler's equations for free surface potential flow.

\subsection{\it Equations of motion}

In what follows we denote the horizontal and vertical components of 
$ {\cal G}^\epsilon$ by 
${\cal G}_1^{\epsilon}$, ${\cal G}_2^{\epsilon}$ respectively.

Then the equations of motion for gravity water waves are 
\begin{equation}
\label{eq-surf-transport}
\partial_t {\cal G}^{\epsilon}(s,t) = 
\nabla \phi \left({\cal G}^{\epsilon}(s,t),t \right), 
\end{equation}   
\begin{equation}
\partial_t \phi(z, t) + 
\frac{1}{2} |\nabla \phi (z,t)|^2 + 
g (y - B) = 0,
\quad\hbox{at}\quad z = [x,y] = {\cal G}^{\epsilon}(s,t), \quad s \in [-h,h]. 
\label{bernouli-1}
\end{equation}
The first equation states that the free surface is transported by the 
flow. The second equation is the Bernoulli equation at the free surface, 
assuming a constant vertical gravitational field of magnitude $g$
and vanishing pressure at the
surface. 

The general idea is that \eqref{eq-surf-transport}, \eqref{bernouli-1}
should imply an evolution equation for the variables $V$, $\phi$, 
equivalently the 
coefficients $a_n$, $b_n$ above. 
This is not obvious since 
$V$ appears implicitly 
in equations \eqref{eq-surf-transport}, \eqref{bernouli-1}.
To obtain  
these evolution equations we use a small amplitude approximation of ${\cal G}^\epsilon $ in terms of $V$.
The $\Phi_n$ will be determined in the process.

\subsection{ \it Explicit approximate representations of the free surface}

The free surface ${\cal G}^{\epsilon} $ can be approximated by its Taylor series at $\epsilon = 0$,
    under standard regularity assumptions on $\nabla V$.
    In what follows we will use the lowest order nontrivial approximation  
\begin{equation}
\label{surface-Euler}
{\cal G}^{\epsilon} (s,t) = [s,B] + \epsilon[\partial_x V([s,B],t),  \partial_y V([s,B],t)] + O(\epsilon^2). 
\end{equation}
Truncating to $O(\epsilon^2)$ we have approximate integration of \eqref{surface-flow} by Euler's rule.
We denote this approximation by $ {\cal G}^\epsilon_E$, and we have 
  \begin{eqnarray}
\label{def-GE}
    {\cal G}^{\epsilon}_E (s,t) & = & [s + \epsilon \partial_x V([s,B],t),
      B + \epsilon \partial_y V([s,B],t)] \\
      \label{GE-surface-approx}
     & = &  
    [s + \epsilon \sum_{n=1}^{\infty}b_n(t) \partial_x \Phi_n([s,B]),
    B +  \epsilon \sum_{n=1}^{\infty}b_n(t) \partial_y \Phi_n([s,B])], 
\end{eqnarray}
by \eqref{psi-expansion}\textcolor{ForestGreen}{.}

We may interpret ${\cal G}^\epsilon_E$ as a small amplitude approximation of 
${\cal G}^\epsilon$. 
By the assumption $ {\hat n} \cdot {\nabla V}(\cdot,t)= 0$ at $\partial W$, 
$\nabla V$ at 
the endpoints $[\pm h, B]$ is parallel to the boundary, 
and the points $[\pm h,B]$ are mapped to points in $\partial W$ by $ {\cal G}^{\epsilon}_E$. The approximation of the surface 
by Euler's rule is therefore quite appropriate for inclined walls of constant slope (assuming that the triple interface 
moves in the regions of constant slope at all times). On the other hand, $ {\cal G}^{\epsilon}_E$
preserves volume automatically, i.e. independently of the dynamics,  
up to $O(\epsilon^2)$, see Proposition \ref{approximate-volume-conserv} and the 
discussion at the end of this section.


As an example of the use of this approximation for ${\cal G}^\epsilon$,  
the velocity at the free surface is 
\begin{equation}
\nabla \phi \left( {\cal G}^{\epsilon}(s,t), t\right) = 
\sum_{n=1}^{\infty}a_n(t)
    [\partial_x \Phi_n\left( {\cal G}^{\epsilon}(s,t), t \right), 
\partial_y \Phi_n\left( {\cal G}^{\epsilon}(s,t), t \right)]
\end{equation}
by \eqref{phi-expansion}. 
By 
\eqref{surface-Euler}   
we have to lowest order  
\begin{equation}
\label{zeroth-order-phi}
\nabla \phi \left( {\cal G}^{\epsilon}(s,t), t \right) = \nabla \phi \left( [s,B] + O(\epsilon) \right) = 
\nabla \phi ( [s,B],t)  + O(\epsilon). 
\end{equation} 

The derivation of approximate models for the dynamical variables proceeds in two steps. 
In the first step we show that the lowest order approximations for
${\cal G}^{\epsilon}$ and $\phi$, and a scaling argument,  
lead to the well-known linear system for inclined wall domains.

\subsection{\it Derivation and analysis of linearized system}

We shall use ${\cal G}^{\epsilon} = {\cal G}^{\epsilon}_E$,
and $\phi = \epsilon {\tilde \phi}$.
Then \eqref{eq-surf-transport}
becomes 
\begin{equation}
\label{linearized-transport}
\epsilon \partial_t [\partial_x V([s,B],t), \partial_y V([s,B],t)] = 
\epsilon [ \partial_x {\tilde \phi} ([s,B],t), \partial_y { \tilde \phi}([s,B],t)] + O(\epsilon^2),
\end{equation}
while \eqref{bernouli-1}
becomes  
\begin{equation}
\label{linearized-bern}
\epsilon \partial_t {\tilde \phi} ([s,B],t) + g \epsilon \partial_y V([s,B],t)  + O(\epsilon^2) = 0, 
\end{equation}
i.e. the nonlinear part is $O(\epsilon^2)$. 
To $O(\epsilon^2)$, 
the second component of the first equation and the second equation 
lead to 
\begin{equation}
\label{lin-evolution}
\partial_{tt} {\tilde \phi} ([s,B],t)  + g \partial_y { \tilde \phi}([s,B],t) = 0. 
\end{equation}

Equation \eqref{lin-evolution}, and 
\begin{equation}
\label{harmonic-cond-tilde}
\Delta {\tilde \phi} = 0 \quad\hbox{in}\quad {\cal W}, \quad
{\hat n} \cdot {\tilde \phi} = 0 \quad\hbox{at}\quad \partial {\cal W},
\end{equation}
form a linear evolution system for ${\tilde \phi}$.  
The first component of \eqref{linearized-transport}
can be used to obtain $V([s,B],t)$ from $\phi([s,B],t)$,
and determine the free surface. 

The $\Phi_n$  
will be the normal modes of
the linearized problem for $\tilde \phi$, that is 
solutions of \eqref{lin-evolution}, \eqref{harmonic-cond-tilde} of 
the form  
\begin{equation}
\label{mode-ansatz}
  {\tilde \phi} ([x,y],t) = \psi([x,y])e^{i\omega t}. 
\end{equation}
Then $\psi $ must satisfy
\begin{equation}
\label{eigenvalue-problem}
\omega^2 \psi ([s,B],t)  = g
\partial_y  \psi([s,B],t),  
\end{equation}
and
\begin{equation}
\label{harmonic-cond}
\Delta {\psi} = 0   \quad\hbox{in}\quad {\cal W},
\quad {\hat n} \cdot \nabla \psi = 0  \quad\hbox{at}\quad \partial {\cal W}.
\end{equation} 

Semi-explicit solutions 
were obtained by Kirchhoff \cite{K79}, see also Lamb \cite{L32},
\textit{Art. 261}.   
We summarize the results. 

The symmetric modes are $\psi = \Psi_S$, where 
\begin{equation}
\label{SymModes}
\Psi_S(x,y;\alpha) = {\cal C}_S(\alpha)[  \cosh(\alpha x)\cos(\alpha y) + \cos(\alpha x)\cosh(\alpha y) ],  
\end{equation}
with ${\cal C}_S(\alpha) $ an arbitrary real. 
The $\Psi_S$ are harmonic in $\mathbb{R}^2$, satisfy  the Neumann boundary condition 
at $\partial {\cal W}$, and are even in $x$, $\forall y \in \mathbb{R}$,  
for all real $\alpha$. By \eqref{eigenvalue-problem}
$\alpha$ must satisfy  
\begin{equation}
\label{sym-cond}
\tanh(\alpha B) = - \tan(\alpha B), 
\end{equation}
and $\omega = \omega(\alpha)$ are given by
\begin{equation}
\label{sym-dispersion}
\omega^2  = g \alpha \tanh(\alpha B). 
\end{equation}

The antisymmetric modes are $\psi = \Psi_A$, where 
\begin{equation}
\label{AsymModes}
\Psi_A(x,y;\alpha)=  {\cal C}_A(\alpha) [\sinh(\alpha x) \sin(\alpha y)+ \sin(\alpha x) \sinh(\alpha y)],
\end{equation}
with $ {\cal C}_A$ an arbitrary real. 
The $\Psi_{A}$ are harmonic in $\mathbb{R}^2$, satisfy  the Neumann boundary condition 
at $\partial {\cal W}$, and are odd in $x$, $\forall y \in \mathbb{R}$,  
for all real $\alpha$. By \eqref{eigenvalue-problem}
$\alpha$ must satisfy  
\begin{equation}
\label{antisym-cond}
 \tanh(\alpha B)= \tan(\alpha B),  
\end{equation}
and $\omega = \omega(\alpha)$ are given by 
\begin{equation}
\label{antisym-dispersion}
\omega^2  = g \alpha \coth(\alpha B). 
\end{equation}

The constants $ {\cal C}_S(\alpha)$, ${\cal C}_A(\alpha_n)$ are arbitrary,  
they will be chosen below to normalize the modes.

There is an additional antisymmetric solution of \eqref{eigenvalue-problem}, see \cite{K79,L32}, that does have the form 
\eqref{AsymModes}, and that we denote by $\Phi_1$, it is given by 
\begin{equation}
\label{lowest-mode}
\Phi_1(x,y) = \sqrt{\frac{3}{2 \pi^5}} xy.  
\end{equation}
By \eqref{eigenvalue-problem} the corresponding frequency $\omega_1$ is given by $\omega^2_1 = g/B$.
We do not consider the zero eigenvalue of 
\eqref{eigenvalue-problem}, it corresponds to the constant harmonic function and does not contribute 
to the expansions for $V$, $\phi$.

The solutions of \eqref{sym-dispersion} (even mode wavenumbers) 
are denoted by $\alpha_{2j+2}$, $j \in \mathbb{N} \cup \{0\}$,  
and satisfy 
\begin{equation}
\label{sym-mode-alpha}
\alpha_{2j+2} B\in (\pi/2+ j \pi, \pi + j \pi), \quad j = 0, 1, 2, \ldots.
\end{equation} 
In the limit $j \rightarrow \infty $ we have $\alpha_{2j + 2}B \rightarrow  j \pi  + 3 \pi/4$, the solutions 
of $- \tan \alpha B = 1$.  

The solutions of \eqref{antisym-dispersion} (odd mode wavenumbers)  
are denoted by $\alpha_{2k+1}$, $k \in \mathbb{Z}^+$,   
and satisfy 
\begin{equation}
\label{antisym-mode-alpha}
\alpha_{2k+1} B\in (k \pi, {\pi}/2 + k \pi), \quad k = 1, 2, \ldots.
\end{equation} 
In the limit $k \rightarrow \infty $ we have $\alpha_{2k+1} B\rightarrow  k \pi +\pi/4$, the solutions 
of $ \tan \alpha B = 1$. 

Combining with \eqref{sym-dispersion}, \eqref{antisym-dispersion}, the dispersion relation
is 
\begin{equation}
\label{disp-relation}
\omega^2_1 = \frac{g}{B}, \quad
\omega_n^2 = 
\begin{cases}
g \alpha_n \tanh {\alpha}_n B, \quad n = 2, 4, 6, \ldots; \\
g  \alpha_n \coth {\alpha}_n B, \quad n = 3, 5, 7, \ldots.
\end{cases}
\end{equation}

\begin{proposition}
\label{Prop-dispersion}
The dispersion relation $\omega_n$, $n \in \mathbb{N}$ of \eqref{disp-relation} is strictly increasing.
\end{proposition}
  
A proof is given in the Appendix. 
We examine the dispersion relation in some more detail in the next section. 
The $\Phi_n$ of \eqref{psi-expansion}, \eqref{phi-expansion} 
are then defined by 
\begin{equation}
\label{def-HA-psi-n}
\Phi_1(x,y) = \Psi_1(x,y), \quad
\Phi_n(x,y) = 
\begin{cases}
\Psi_S(x,y;\alpha_n), \quad n = 2, 4, 6, \ldots; \\
\Psi_A(x,y;\alpha_n), \quad n = 3, 5, 7, \ldots. 
\end{cases}
\end{equation}

We also let 
\begin{equation}
\label{def-ksi}
\xi_n(x) = \Phi_n(x,B), \quad n \in \mathbb{N},  
\end{equation}

The real numbers  $ {\cal C}_S(\alpha)$, ${\cal C}_A(\alpha)$, defined for $\alpha = \alpha_n$, $n$ odd and even 
positive integers 
respectively are still  
arbitrary.  We use the notation $ {\cal C}_S(n) = {\cal C}_S(\alpha_n)$ for $n$ even, and 
 $ {\cal C}_A(n) = {\cal C}_A(\alpha_n)$ for $n \geq 3$ odd. 
 In what follows we choose them so that $\int_{-h}^h |\xi_n|^2  = 1$, $\forall n \mathbb{N}$.
$\xi_1$ is already normalized for $h = \pi$.

By \eqref{eigenvalue-problem},
\eqref{harmonic-cond}
the $\xi_n$ are eigenfunctions of the flat surface Dirichlet-Neumann operator ${\hat \partial}_y$, 
defined by 
\begin{equation}
\label{def-flat-surf-DN}
 {\hat \partial}_y \xi_n(s) = \partial_y \Phi_n(s,B) = g^{-1} \omega^2_n \Phi_n(s,B) = g^{-1} \omega^2_n \xi_n(s),
\end{equation}
for all $s \in [-h,h]$, and $n \in \mathbb{Z}^+$. 

Let $ X = L^2([-h,h];\mathbb{R})$ with the standard inner product $\langle f,g \rangle  = \int_{-h}^{h} f(s) g(s)\; ds$. 
The operator ${\hat \partial}_y$ is a symmetric operator in $X$
with dense domain, 
moreover the 
$\{\xi_n \}_{n= 1}^{\infty}$ form a orthonormal set in $X$.
Orthogonality follows by Proposition \ref{Prop-dispersion} and Green's identity. 
Also, \eqref{def-flat-surf-DN}, $\omega_n \neq 0$, implies that the $\xi_n$ have zero average, $\forall n \geq 1$. 
Some related spectral theory is presented in \cite{G94} for the related problem on channels with constant cross section.

\subsection{ \it Derivation of weakly nonlinear model}
 
In the second step of deriving approximate equations for the dynamical variables  
we use the approximation of ${\cal G}^{\epsilon}$ by ${\cal G}^{\epsilon}_E$
to write an approximate evolution equation that includes the linearized equation
above and a lowest order nonlinear part. 

We use 
${\cal G}^{\epsilon} = {\cal G}^{\epsilon}_E$, ${\cal G}^{\epsilon}_E$ as in  
\eqref{def-GE}, and $\phi = \epsilon \psi$, then
the first equation \eqref{eq-surf-transport} is  
\begin{eqnarray}
\label{first-transp}
\epsilon \partial_t \partial_x V([s,B],t) & = & \epsilon \partial_x \psi(s + \epsilon \partial_x V([s,B],t), 
B + \epsilon \partial_y V([s,B],t),t), \\
\label{second-transp}
\epsilon \partial_t \partial_y V([s,B],t) & = & \epsilon \partial_y \psi(s + \epsilon \partial_x V([s,B],t), 
B + \epsilon \partial_y V([s,B],t),t).
\end{eqnarray}

Expanding the right-hand side we have
\begin{eqnarray}
\nonumber
\partial_t \partial_x V ([s,B],t) & = & \partial_x \psi([s,B],t)  + \epsilon
\Big[ (\partial^2_{x} \psi([s,B],t) ) \partial_x V([s,B],t)  + \\
\label{expand-transp-first}
&  + & 
(\partial_y \partial_{x} \psi([s,B],t) ) \partial_y V([s,B],t)  
\Big]  + O(\epsilon^2), 
\end{eqnarray}
\begin{eqnarray}
\nonumber
\partial_t \partial_y V ([s,B],t)  & = & \partial_y \psi ([s,B],t) + 
\epsilon \Big[ (\partial_{x}\partial_y \psi ([s,B],t)) \partial_x V ([s,B],t)) +
\\
\label{expand-transp-second}
& + & 
(\partial^2_{y} \psi ([s,B],t))) \partial_y V ([s,B],t))
\Big]  + O(\epsilon^2). 
\end{eqnarray}
The $\epsilon-$term can be abbreviated as 
$ (\nabla V([s,B],t) ) \cdot \nabla \partial_y \psi([s,B],t) $.

Similarly, the Bernoulli equation \eqref{bernouli-1} becomes
\begin{eqnarray}
\nonumber
\epsilon \partial_t \psi([s,B]  & + & \epsilon (\nabla \partial_t \psi([s,B],t))) \cdot \nabla V([s,B],t), t)  + \\ 
\label{expand-bernouli}
& + & \frac{1}{2} \epsilon^2| \nabla \psi([s,B] + \epsilon \nabla V([s,B],t), t)|^2 + 
g \epsilon \partial_y V([s,B],t) = 0, 
\end{eqnarray}
therefore 
\begin{eqnarray}
\nonumber
\partial_t \psi ([s,B],t)  & +  & g \partial_y V([s,B],t) + 
\epsilon 
\Big[ \nabla \partial_t \psi([s,B],t))\cdot \nabla V([s,B],t) +  \\
\label{approx-bernouli}
& + &  
\frac{1}{2} |\nabla \psi([s,B],t)|^2 
\Big] = O(\epsilon^2).
\end{eqnarray} 

The first approximate  nonlinear model is obtained by 
omitting the $O(\epsilon^2)$ terms in 
\eqref{expand-transp-first},
\eqref{expand-transp-second},
\eqref{approx-bernouli}.

The dynamical variables are $\psi(\cdot,t)$ and $V(\cdot,t)$, equivalently the coefficients
$a_n(t)$, $b_n(t)$, $n \in \mathbb{N}$, of  \eqref{phi-expansion}, \eqref{psi-expansion}
respectively. 
To obtain the approximate evolution equations for $a_n$, $b_n$,  
we will consider a system of two equations, one of the approximate 
transport equations 
\eqref{expand-transp-first},
\eqref{expand-transp-second}, 
and the approximate Bernoulli \eqref{approx-bernouli}.

Note that \eqref{approx-bernouli}
is implicit for $\partial_t \psi$. We 
have a term 
\begin{equation}
\label{approx-bernouli2}
\partial_t \psi + \epsilon 
(\partial_t \nabla \psi)\cdot \nabla V = 
\sum_{n=1}^\infty {\dot a}_n(t) \left(  \Phi_n(s,B)  +  \epsilon \nabla \Phi_n(s,B) \cdot \nabla V([s,B]), t) \right).   
\end{equation}
We also need to expand $\nabla V$, so that we will have 
sums of terms $ {\dot a}_n b_m$, and we need to invert an operator to write the 
equation for the ${\dot a}_n$ and $\partial_t \psi$.  

One way to avoid this complication is to write 
$\partial_t \nabla \psi = \nabla \partial_t \psi $, and use 
$$  \partial_t \psi([s,B],t)= - g \partial_y V([s,B],t)  + O(\epsilon) $$
from \eqref{approx-bernouli}.
Then the approximate Bernoulli equation 
is, up to $O(\epsilon^2)$, 
\begin{eqnarray}
\nonumber
\partial_t \psi ([s,B],t)  & = & -  g \partial_y V([s,B],t) + \\
\label{trunc-approx-bernouli}
& + & 
\epsilon 
\Big[ g (\nabla \partial_y V([s,B],t)) \cdot \nabla V([s,B],t) - 
\frac{1}{2} |\nabla \psi([s,B],t)|^2 \Big]. 
\end{eqnarray}

Using the vertical component of the approximate transport equation
\eqref{expand-transp-second} to $O(\epsilon^2)$,  
and the approximate Bernoulli equation \eqref{trunc-approx-bernouli}
we obtain the approximate model  
\begin{eqnarray}
\nonumber
\partial_t \partial_y V ([s,B],t) & = & \partial_y \psi([s,B],t) + 
\epsilon \Big[ (\partial_{x} \partial_y \psi ([s,B],t) ) \partial_x V ([s,B],t)+
\\
\label{first-approx-syst}
& + & 
(\partial^2_{y} \psi ([s,B],t)) \partial_y V ([s,B],t)
\Big] , \\
\nonumber
%
\nonumber
\partial_t \psi ([s,B],t) & = & -  g \partial_y V ([s,B],t)+  
\epsilon 
\Big[  g (\nabla \partial_y V([s,B],t)) \cdot \nabla V ([s,B],t)+ 
\\
\label{second-approx-syst}
& - & 
\frac{1}{2} |\nabla \psi([s,B],t)|^2 \Big]. 
\end{eqnarray}

\begin{remark}
\label{implicit-psi-evol}
Using $\partial_t \nabla \psi = \nabla \partial_t \psi $, 
the left-hand side of 
{\eqref{approx-bernouli2}}is 
$$ (1 + \epsilon (\nabla V([s,B],t)) \cdot \nabla )\partial_t\psi(\textcolor{ForestGreen}{[}s,B\textcolor{ForestGreen}{]},t). $$ 
We can make the equation explicit using the formal 
inverse 
$$ (1 + \epsilon (\nabla V([s,B],t)) \cdot \nabla)^{-1} 
= 1 -   \epsilon (\nabla V([s,B],t)) \cdot \nabla + O(\epsilon^2). $$
The result is again \eqref{trunc-approx-bernouli}.
The inversion should make sense provided we have fast decay of the coefficients $b_n$. 
A similar inversion appears in the Haut-Ablowitz method for computing the 
Dirichlet-Neumann operator
\cite{AH08,WV15}; this step therefore reflects some issues related to the Dirichlet-Neumann operator. 
\end{remark}

An alternative model is obtained using the horizontal component 
\eqref{expand-transp-first}
of the approximate transport equation to $O(\epsilon^2)$, 
and \eqref{trunc-approx-bernouli}. The main reason for choosing to continue using the equation for 
the vertical component is the analogy of the variable $\partial_y V$ with a wave height. Otherwise, 
both models should be considered as equivalent necessary conditions, and their relation should be examined 
in more detail.

\subsection{\it Spectral representation of model equation and generalizations}

We now write the spectral form of equations 
\eqref{first-approx-syst},
\eqref{second-approx-syst}.
We insert expansions \eqref{psi-expansion}, \eqref{phi-expansion} into 
\eqref{first-approx-syst},
\eqref{second-approx-syst}.
The equations involve compositions of partial derivatives of  $V$, $\psi$, 
evaluated at $[x,y]= [s,B]$.
By linearity, 
the partial derivatives  
are applied to functions $\Phi_n(x,y)$, and evaluated at $[x,y]= [s,B]$. Clearly 
$\partial_x$, $\partial_y$ commute, and all partial derivatives  
are continuous at $[s,B]$, $s \in [-h,h]$. 

Definitions 
\eqref{def-HA-psi-n}, \eqref{def-ksi}
for 
$\Phi_n$, $\xi_n$, respectively, $n \geq 2$, imply  
\begin{eqnarray}
\nonumber
\partial_x \Phi_n(s,B) & = & \xi'_n(s), \\
\nonumber
\partial_y \Phi_n(s,B) & = & g^{-1}\omega_n^2 \Phi_n(s,B) = g^{-1}\omega_n^2 \xi_n(s), \\
\nonumber
\partial^2_x \Phi_n(s,B) & = & \xi''_n(s), \\
\nonumber
\partial_y \partial_x \Phi_n(s,B) & = & \partial_x \partial_y \Phi_n(s,B) = g^{-1}\omega^2_n \partial_x \Phi_n(s,B) =  
g^{-1}\omega^2_n \xi'_n(s) \\
\label{surface-operations}
\partial^2_y \Phi_n(s,B) & = &  -  \partial^2_x \Phi_n(s,B) = - \xi''_n(s).
\end{eqnarray}
The above also hold for $\Phi_1$.

We then multiply \eqref{first-approx-syst},
\eqref{second-approx-syst}
by $\xi_m$,  
integrate over $s \in [-h,h]$, and use the orthogonality of the $\xi_n$. 
We also use the fact that the $\xi_n$ are normalized. 

Then equation \eqref{first-approx-syst} leads to 
\begin{equation}
\label{first-spectral-eq}
{\dot b}_m =  {a }_m + 
\epsilon \omega_m^{-2}
\sum_{n_1,n_2 = 1}^\infty 
\left( \omega^2_{n_1}  I_{n'_1, n'_2,m}
-  \omega^2_{n_2} I_{n''_1, n_2,m} \right) a_{n_1} b_{n_2}, 
\quad m \in \mathbb{N},
\end{equation}
with  
\begin{equation}
\label{spectral-coef-1}
I_{n'_1, n'_2,m} = 
\int_{-h}^{h} \xi'_{n_1}(s) \xi'_{n_2}(s) \xi_m(s) \; ds, \quad 
I_{n''_1, n_2,m} = 
\int_{-h}^{h} \xi''_{n_1}(s) \xi_{n_2}(s) \xi_m(s) \; ds.
\end{equation}

Similarly, equation \eqref{second-approx-syst} leads to  
\begin{eqnarray}
\nonumber
{\dot a}_m & = &  - \omega_m^2 {b }_m 
+ \epsilon  
\sum_{n_1,n_2 = 1}^\infty
\left( \omega^2_{n_1} I_{n'_1, n'_2,m} 
-
 \omega^2_{n_2}I_{n''_1, n_2,m} \right) b_{n_1} b_{n_2} \\
\label{second-spectral-eq}
& &  -  
\frac{\epsilon}{2} 
\sum_{n_1,n_2 = 1}^\infty 
\left( I_{n'_1, n'_2,m} 
+ 
g^{-2} \omega^2_{n_1} \omega^2_{n_2} I_{n_1, n_2,m} \right) a_{n_1} a_{n_2}
, \quad m \in \mathbb{N},
\end{eqnarray}
with $I_{n'_1, n'_2,m} $, $I_{n_1, n_2,m}$ 
as in \eqref{spectral-coef-1}, and 
\begin{equation}
\label{spectral-coef-2}
I_{n_1, n_2,m} = 
\int_{-h}^{h} \xi_{n_1}(s) \xi_{n_2}(s) \xi_m(s) \; ds.
\end{equation}

The coefficients $I_{n_1'',n_2,n_3}$, $I_{n'_1, n'_2,m} $, $I_{n_1, n_2,m}$ can be computed explicitly
since 
the $\xi_n$ and their derivatives are trigonometric and hyperbolic functions. 
We discuss their computation and structure in the next section.

\begin{remark} 
\label{even-subspace}
We observe that $\psi$, $V$ even imply that the right hand sides of 
\eqref{first-approx-syst}, \eqref{second-approx-syst} are even. 
We use the fact 
that $ {\hat \partial}_y$ preserves parity.
Therefore 
the subspace of $\psi$, and $V$ both even is invariant and  
we may study the expansion with even integer indices.  
These observations also follow by examining the 
coefficients $I_{n_1'',n_2}$, $I_{n'_1, n'_2,m} $, $I_{n_1, n_2,m}$, e.g. they 
vanish for $m$ even and odd $n_1$, $n_2$.
Other combinations of parities for 
$\psi$, $V$ do not lead to invariant subspaces.
\end{remark}

The computation of the spectral equations from 
\eqref{first-approx-syst},
\eqref{second-approx-syst} 
suggests writing system   
\eqref{first-approx-syst},
\eqref{second-approx-syst} as an evolution equation for quantities on the 
interval $[-h,h]$. Formulas \eqref{surface-operations}
and generalizations for derivatives of arbitrary order
map of expressions partial derivatives $\partial_x$, $\partial_y$
of harmonic functions, evaluated at points $[x,y] = [s,B]$, $s \in [-h,h]$, to expressions involving the 
horizontal derivative $\partial_s$ and the operator ${\hat \partial}_y$ 
defined in \eqref{def-flat-surf-DN}. 
The map is extended linear combinations of the 
$\Phi_n$ by linearity. 
Letting 
$$  \zeta(s,t) = \psi(s,B,t), \quad \beta(s,t) = V(s,B,t), $$
\eqref{first-approx-syst},
\eqref{second-approx-syst}, and 
\eqref{surface-operations}, \eqref{def-flat-surf-DN} are written as  
\begin{eqnarray}
\label{surface-eq-1}
\partial_t  {\hat \partial_y} \beta  & = & {\hat \partial}_y \zeta 
 + \epsilon \Big[ ( \partial_s {\hat \partial}_y \zeta  ) \partial_s \beta   
- ( \partial^2_{s} \psi ) {\hat \partial}_y \beta \Big] , \\
\label{surface-eq-2}
\partial_t \zeta & = & -  g {\hat \partial}_y \beta + 
\epsilon 
\Big[  g \left( ( \partial_s {\hat \partial}_y  \beta )  \partial_s \beta  - (\partial^2_s \beta ) {\hat \partial}_y \beta \right) 
+  \frac{1}{2} \left( (\partial_s \zeta )^2 +  ({\hat \partial}_y \zeta)^2 \right)  \Big]. 
\end{eqnarray}

Note that $\beta$, $\zeta $ satisfy $(\pm \partial_s - {\hat \partial}_y)\beta = 0 $ at $s = \pm h$, and 
$(\pm \partial_s - {\hat \partial}_y)\zeta = 0 $ at $s = \pm h$, at all times $t$. 
We expect that these conditions at the boundary are implied by 
the definition of ${\hat \partial}_y$.
The free surface at time $t$ is (defined as) the set of points 
\begin{equation}
\label{free-surf-def}
[s + \epsilon \partial_s \beta(s,t),
      B  +  \epsilon {\hat \partial}_y \beta(s,t)], \quad s \in [-h,h]. 
\end{equation}
This definition is analogous to 
\eqref{def-GE}, but we may alternatively use an expression obtained from  
a higher order Taylor expansion of ${\cal G}^\epsilon$ with 
horizontal and vertical partial derivative operators replaced by 
suitably ordered operators $\partial_s$, ${\hat \partial}_y $. 
The ordering is obtained by a generalization of the correspondence    
\eqref{surface-operations} to higher derivatives. The general formula is easily obtained and is omitted.


System  \eqref{surface-eq-1}, 
\eqref{surface-eq-2} may be useful in alternative numerical discretizations 
and approximations of the operators $\partial_s$, ${\hat \partial}_y $. Also 
we may consider \eqref{surface-eq-1}, 
\eqref{surface-eq-2} in more general domains. 
Elliptic equations on domains with Lipschitz boundary have been studied extensively \cite{G11}, 
and the continuity of the normal derivative at corners, e.g. the endpoints of the free surface, is not guaranteed even for smooth Dirichlet data. 
The results of \cite{MW16} suggest that the solution of Laplace's equations in domains such as polygons can have an $H^2$ component 
that is localized at the edges and 
whose regularity can not be improved by considering smoother Dirichlet data at the free surface.   
This issue is not apparent in our construction. The first ingredient of our approach is a family of 
functions that satisfy the boundary condition and are harmonic in a larger domain. 
The computation of the Dirichlet-Neumann 
operator for linear combinations of such functions involves computation of the coefficients of the series for a given
Dirichlet boundary condition, 
and term-wise 
application of $\partial_{\hat n}$ to the sum \cite{AH08}. 
The first step is more challenging, see \cite{WV15}) and Remark \ref{implicit-psi-evol}, but the procedure 
also suggests ways to control the regularity of the normal derivative. 
A second ingredient of our approach are expansions in eigenfunctions of the Dirichlet-Neumann operator. 
These functions can have additional regularity and the normal derivative can be controlled by the expansion 
coefficients of the boundary condition. Both ingredients are here related and we also expect that they 
are available in more general domains. 
For instance, we can produce a large class of harmonic functions in a bounded domain $\cal W$ that satisfy
the Neumann condition at a subset $\Gamma_1  \subset {\partial {\cal W}}$ by considering a larger domain 
${\tilde {\cal W}} \supset {\cal W}$, 
and solving Laplace's equation 
in $\tilde {\cal W}$ with 
Neumann condition in ${\tilde \Gamma}_1 \subset {\partial {\cal W}}$, ${\tilde \Gamma}_1 \supset \Gamma_1$,
and Dirichlet conditions in 
$\partial {\cal W} \setminus {\tilde \Gamma}_1$. 
Even in the case where $\tilde {\cal W}$ has corners at the intersection of 
the Dirichlet and Neumann boundaries, the normal derivative at the free surface, assumed to lie in $\cal W$, 
is applied at points where the regularity of the 
solution can be made arbitrarily high, i.e. the free surface of $\cal W$ consists of points that are either in the interior of $\cal W$ or the 
smooth part of $\partial W$, see \cite{MW16}, Prop. 5.19.  
The interpretation of \eqref{surface-eq-1}, \eqref{surface-eq-2} 
as a reasonable model for general domains will require further development of the expansion ideas and comparison with the 
elliptic theory.
The next section suggests that while  
expansions may make \eqref{surface-eq-1}, \eqref{surface-eq-2}
meaningful for more general domains,  
effective computation requires alternative schemes to approximate the Dirichlet-Neumann operator.

\subsection{\it Further remarks on surface parametrization}

We conclude this section with some further remarks on the proposed representation of the free surface, and 
the derivation of the approximate model. 

The fact that $ {\nabla V}$ is parallel to $\partial W$
allows us to use the expression defining $ {\cal G}^\epsilon_E$
as an alternative exact representation of the free surface, see \eqref{free-surf-def}. 
This alternative representation can be used only for constant slope beaches.
Also the expression defining $ {\cal G}^\epsilon_E$
leads to automatic mass conservation, i.e. independently of the
dynamics, only up to an $O(\epsilon^2)$. 
Analogous results are expected for higher order Taylor approximations 
of ${\cal G}^\epsilon$.

\begin{proposition}
\label{approximate-volume-conserv}
Let 
$$[X(s), Y(s)] = [s + \epsilon \partial_x V(s,B), B + \epsilon \partial_y V(s,B)],$$ 
$s \in [-h,h]$, $\epsilon \in \mathbb{R}$, 
with $V$ harmonic in $\cal W$, and ${\hat n} \cdot \nabla V = 0$ at $\partial {\cal W}$.
Assume also that the image of $[X,Y]$ intersects 
$\partial {\cal W}$ only at the endpoints $s=\pm h$.
Then 
  \begin{equation}\label{gradient-vol-conserv}
 \frac{1}{2}  \int_{-h}^{h}(Y(s)\;dX - X(s)\;dY)  =  B^2 + O(\epsilon^2).
\end{equation}
\end{proposition}

The proof is in the Appendix. 



\begin{remark}
The coefficients $b_n(t)$ of $V$ obtained using the approximate system   
\eqref{first-spectral-eq}, \eqref{second-spectral-eq}, 
can be used to calculate 
the free surface by integrating \eqref{surface-flow} in $\tau$ at each $t$. More generally
we may use   
approximations of different order of ${\cal G}^{\epsilon}$ in approximating the evolution
equations and in computing of the free surface at each time. 
\end{remark}

We have not examined however the use of higher order expansions of ${\cal G}^{\epsilon}$ in deriving 
model equations. Such expansions may have several drawbacks, e.g. they will involve more derivatives. 
As we see in the next section and the Supplement there are also practical limitations, e.g. computations of 
mode interactions.

\begin{remark}
The description of the free surface 
by ${\cal G}^{\epsilon}$, or by approximations such 
as ${\cal G}_E^{\epsilon}$ is that we can have criteria 
for the surface to be the graph of a function. 
Clearly, if the map $x \mapsto X$ defined by 
$X(x,t)$, $x \in [-h,h]$,  
is invertible, then we may use the inverse $x(X,t) $ to write 
the vertical component $Y$ as  
$Y(x,t) = Y(x(X,t),t)$, i.e. a function of the horizontal component $X$, 
$X \in [X(-h,t),X(h,t)]$. 
For instance, invertibility of $X(x,t) = x + \epsilon \partial_x V(x,t)$
would follow from $ || \partial_x V(\cdot,t)||_{C^1} \leq M$ and $|\epsilon|$
sufficiently small.

In the case where $[X(\cdot,t),Y(\cdot,t)]$ is a 
graph but is  
given by an approximation 
of $ {\cal G}^{\epsilon}$ we may also want to 
prove a lower bound on $Y(X(\cdot,t),t)$ that implies that 
the surface does not intersect $\partial {\cal W}$. 
(A bound on the derivative on $dY/dX$ would be sufficient.)
\end{remark}

%

A limitation of the theory is that  
the generality of the curves obtained by the proposed potential description \eqref{surface-flow}, \eqref{psi-expansion} 
of the free surface is not known at the moment. 
The fact that the surface is described by an infinite set of parameters, the coefficients $b_n$, 
and the recovery of the linear theory suggest sufficient generality, but a    
more conclusive statement is left for further work. 

\section{Numerical study of low mode interactions}

We consider now some truncations of the spectral equations and their numerical integration.
The goal is to examine some simple models involving the lowest modes
and indicate some computations with the simplified model.
Before presenting the simulations we discuss the dispersion relation and
some basic features  of  
spectral system  \eqref{first-spectral-eq}, 
\eqref{second-spectral-eq} and finite mode truncations. Writing these systems involves the 
computation of mode interaction coefficients. These computations are lengthy  
and we present some details in a Supplement.
We restrict our attention to the subspace of symmetric modes.
 
\begin{table}
    \centering
    \begin{tabular}{ccccc}
        \noalign{\smallskip}\hline\noalign{\smallskip}     
   &    &  $B \alpha_n$ &      $ \sqrt{\frac{B}{g}}\omega_n$ &   $\sqrt{B \alpha_n}$  \\ \hline
        \noalign{\smallskip}  \hline\noalign{\smallskip}  
mode 1   & odd 1     &                NA                    & 1                                       &      NA                               \\ \hline
mode 2  & even 1      & 2.365020372431352  & 1.5243483044999009 & 1.5378622735574703 \\ \hline  
mode 3  & odd 2      & 3.926602312047919   & 1.9823356191069623 & 1.9815656214336983 \\ \hline
mode 4  & even 2      & 5.497803919000836  & 2.3447002939328567 & 2.3447396271229852 \\ \hline
mode 5  & odd 3     & 7.068582745628732   & 2.658682567428893  & 2.658680640022177 \\ \hline
mode 6  & even 3      & 8.63937982869974   & 2.939282104084953  & 2.9392821961662237 \\ \hline
mode 7  & odd 4       & 10.210176122813031  & 3.1953366255307207 & 3.195336621204882  \\ \hline
mode 8  & even 4      & 11.780972451020228  & 3.4323421230468543 & 3.432342123247656 \\ \hline
        \noalign{\smallskip}\hline
    \end{tabular}
    \caption{
    The first column is the mode index: even for symmetric modes, and odd for antisymmetric modes. 
The second column shows $B \alpha_n$, $n \geq 2$, the solutions of 
\eqref{sym-cond}, 
\eqref{antisym-cond} for $n $ odd, even respectively.
The third column shows the scaled frequencies 
$\sqrt{B/g} \omega_n =  \sqrt{B \alpha_n \tanh B \alpha_n}$  
for $n \geq 2$ even, and 
$ \sqrt{B/g} \omega_n = \sqrt{ B \alpha_n \coth B \alpha_n} $  
for $n \geq 3$ odd. 
The fourth column shows $ \sqrt{B \alpha_n}$, $n \geq 2$, the large$-n$ limit.
}
\label{freq-table} 
\end{table}

\subsection{ \it Dispersion relation and mode interaction coefficients}

We first compute  
the frequencies $\omega_n$, $n \geq2$, of the linear part.   
Numerical values of $\alpha_n$, see \eqref{sym-dispersion}, \eqref{antisym-dispersion}, 
and the resulting 
$\omega_n$, see 
\eqref{disp-relation}, 
are shown in Table \ref{freq-table}.
Equations 
\eqref{sym-dispersion}, \eqref{antisym-dispersion}
for the $\alpha_n$ are solved by Newton's method \cite{R68,Z86}. The iteration for each $n$ is started from 
the large$-n$ asymptotic value $ \frac{\pi}{2}(n - \frac{1}{2})$, $n \geq 2$. 
These values are already good approximations from the smallest $n$, and it can be shown 
that the Newton iteration 
converges for all $n \geq 2$. 
The dispersion relation also approaches rapidly the deep water limit $ \sqrt{B \alpha_n}$, 
$B \alpha_n \sim \frac{\pi}{2}(n - \frac{1}{2})$ and we can also show that 
the error decays exponentially in the mode index. (The analysis of the Newton iteration
will be presented elsewhere.) 
The fact that the dispersion relation is monotonic, Proposition \ref{Prop-dispersion},   
follows from a simpler argument, see Appendix. 


In the numerical simulations below we consider  
small truncations in the symmetric mode subspace, see Remark \ref{even-subspace}, using the lowest frequency modes.
The spectral equations involve the coefficients
\begin{eqnarray}
\nonumber
{\cal A}(n_1,n_2,n_3) & = & \omega^2_{n_1}  I_{n'_1, n'_2,n_3} -  \omega^2_{n_2} I_{n''_1, n_2,n_3}, \\
\label{cubic-A-B-coeff}
{\cal B}(n_1,n_2,n_3) & = & I_{n'_1, n'_2,n_3}  + g^{-2} \omega^2_{n_1} \omega^2_{n_2} I_{n_1, n_2,n_3},  
\end{eqnarray}
with $I_{n_1'',n_2,n_3}$, $I_{n'_1, n'_2,n_3} $, $I_{n_1, n_2,n_3}$ defined by 
\eqref{spectral-coef-1},  \eqref{spectral-coef-2}. The $\xi_n$, $n \geq 1$, are assumed normalized. 
Note that 
some combinations of indices lead to vanishing coefficients by parity considerations:  
for $n_3 $ even, and $n_1$, $n_2$  of opposite parity 
the triple integrals $I_{n_1'',n_2,n_3}$, $I_{n'_1, n'_2,n_3} $, $I_{n_1, n_2,n_3}$  vanish, therefore 
$ {\cal A}(n_1,n_3,n_3) =  {\cal B}(n_1,n_1,n_3) = 0$. 
Also $ {\cal B}(n_2,n_1,n_3) =   {\cal B}(n_1,n_2,n_3) $ for all $n_1$, $n_2$, $n \in \mathbb{N}$.

For instance the truncation of 
\eqref{first-spectral-eq},
\eqref{second-spectral-eq}
to modes with index $n \in \{2,4\}$ is 
\begin{eqnarray}
\label{b-2-eq}
{\dot b}_{2}  & = & a_{2} + \epsilon \omega_2^{-2} [{\cal A}(2,2,2) a_2 b_2 + {\cal A}(2,4,2)a_2 b_4  + {\cal A}(4,2,2)a_4 b_2 + 
  {\cal A}(4,4,2) a_4 b_4 ], \\ 
 \label{b-4-eq}
{\dot b}_4  & = & a_4 + \epsilon  \omega_4^{-2}[{\cal A}(2,2,4) a_2 b_2 + {\cal A}(2,4,4)a_2 b_4  + {\cal A}(4,2,2)a_4 b_2 + 
  {\cal A}(4,4,4) a_4 b_4 ], \\
\label{a-2-eq}
{\dot a}_2  & = &  - \omega_2^2 b_2 + \epsilon [ {\cal A}(2,2,2) b^2_2  + ({\cal A}(2,4,2) + {\cal A}(4,2,2))b_2 b_4  + {\cal A}(4,4,2) b_4^2  ] \\ 
\nonumber
& & - \frac{\epsilon}{2} [  {\cal B}(2,2,2) a_2^2  + 2 {\cal B}(2,4,2) a_2 a_4 + {\cal B}(4,4,2) a_4^2], \\ 
\label{a-4-eq}
{\dot a}_4  & = &  - \omega_4^2 b_4 +\epsilon [    {\cal A}(2,2,4) b^2_2  + ({\cal A}(2,4,4) + {\cal A}(4,2,4))b_2 b_4  + {\cal A}(4,4,4) b_4^2  ] \\ 
\nonumber
& & - \frac{\epsilon}{2}[    {\cal B}(2,2,4) a_2^2  + 2 {\cal B}(2,4,4) a_2 a_4 + {\cal B}(4,4,4) a_4^2 ]. 
\end{eqnarray}
A simpler system describing only one even mode is 
\begin{eqnarray}
\label{one-mode-b-eq}
{\dot b}_n  & = & a_n  + \epsilon \omega_n^{-2}  {\cal A}(n,n,n) a_n b_n, \\ 
\label{one-mode-a-eq}
{\dot a}_n & = &  - \omega_n^2 b_n + \epsilon  {\cal A}(n,n,n) b^2_n  - \frac{\epsilon}{2} {\cal B}(n,n,n) a_n^2, 
\end{eqnarray}
with $n$ even.

\begin{table}
    \centering
    \begin{tabular}{ccccc} \noalign{\smallskip}\hline\noalign{\smallskip}     
$(n_1,n_2,n_3)$& Analytic Expression  & Clenshaw-Curtis  \\ \hline
        \noalign{\smallskip}  \hline\noalign{\smallskip}  
$(2,2,2)$  &  -0.090326909807416 &.  \textbf{-0.09032}5631806652
  \\ 
\hline  
$(2,4,2)$  & 2.338779339988490& \textbf{2.33877}6342495788
       \\ \hline
$(4,2,2)$  & 4.358824845127728&  \textbf{4.3588}17698222419  \\
\hline
$(4,4,2)$  &  -1.863286560769736 &  \textbf{-1.8632}69738054163 &   \\ \hline  
$(2,2,4)$ &-0.171874228552746& \textbf{-0.17187}5506553511 \\ \hline
$(2,4,4)$  &0.575690065541405&    \textbf{0.57569}3086270485   \\
\hline
$(4,2,4)$  & 0.173666012259148&   \textbf{0.1736}73159164457 \\
\hline  
$(4,4,4)$  & 0.488707254211797&  \textbf{0.488}690431496224 \\
        \noalign{\smallskip}\hline
    \end{tabular}
    \caption{Coefficients ${\cal A}(n_1,n_2,n_3)$ of \eqref{cubic-A-B-coeff}. In the first column the triple integrals  
    \eqref{spectral-coef-1}, \eqref{spectral-coef-2} were evaluated analyticaly; in the second the integrals were computed by quadrature.
    We use $B = h = \pi$, $g = 1$.}
    \label{tab:AandB}    
\end{table}

\begin{table}
    \centering
    \begin{tabular}{ccc}
        \noalign{\smallskip}\hline\noalign{\smallskip}     
$(n_1,n_2,n_3)$&  Analytic Expression & Clenshaw-Curtis  \\ \hline
        \noalign{\smallskip}  \hline\noalign{\smallskip}  
$(2,2,2)$   &-1.110537680120724 &  \textbf{-1.1105}20716271800\\ 
\hline  
$(2,4,2)$  &   4.881518557581592&   \textbf{4.881}478305429332 \\ \hline
$(4,2,2)$  & 4.839883447581593& \textbf{4.8398}43185429332  \\
\hline
$(4,4,2)$  &-2.110838537003454& \textbf{-2.110}743505919745 \\ \hline  
$(2,2,4)$ & 1.563850507932040&  \textbf{1.5638}33490090192\\ \hline
$(2,4,4)$  & -0.757159525024813 &   \textbf{-0.7571}19352293886 \\
\hline
$(4,2,4)$  & -0.717331695024813 & \textbf{-0.717}291522293886\\
\hline  
$(4,4,4)$   & 2.783582282653778& \textbf{2.783}487251570069\\
        \noalign{\smallskip}\hline
    \end{tabular}
    \caption{Coefficients  ${\cal B}(n_1,n_2,n_3)$ of \eqref{cubic-A-B-coeff}. In the first column the triple integrals  
    \eqref{spectral-coef-1}, \eqref{spectral-coef-2} were evaluated analytically; in the second the integrals were computed by quadrature.
    We use $B = h = \pi$, $g = 1$.}
    \label{tab:B} 
\end{table}

Before presenting numerical integration results 
we summarize the computation of the interaction coefficients, details are in the  Supplement. 
We first describe the normalization of the $\xi_n$, $n \geq 2$. The mode $\xi_1$ is already normalized. 

To avoid constants that can grow exponentially in space and in the index $n$  
we write the 
symmetric modes as 
\begin{equation}
\label{s1}
{\tilde \xi}_m(s)=   \cos \alpha_m B  \frac{\cosh \alpha_m s}{ \cosh \alpha_m B } +\cos \alpha_m s, 
\end{equation}
$m$ even. 
We have an oscillatory part, and the ``edge'' part 
\begin{equation}
\label{edge-fun}
E_m(s)  =  \frac{\cosh \alpha_m s}{ \cosh \alpha_m B } = 
e^{\alpha_m(s - B)} \frac{1 + e^{- 2 \alpha_m s }}{1 + e^{- 2 \alpha_m B }},
\end{equation}
with values in the interval $[0,1]$. The edge part thus decays exponentially away from $s = \pm h$.

In what follows we consider the case $B = h = \pi$.
The normalization factors   
$$ {\tilde c}_m^2 =   \int_{-\pi}^\pi  | {\tilde \xi}_m(s) |^2 \; ds = 2  \int_{0}^\pi  | {\tilde \xi}_m(s) |^2 \; ds,  $$
for $m \geq 2$, even, are then  
\begin{equation}
\label{normalization-form}
{\tilde c}_m^2  = \pi + \frac{\sin 2 \alpha_m \pi }{2 \alpha_m}
+  \frac{ \cos^2 \alpha_m \pi (\sinh 2 \alpha_m \pi  + 2 \pi \alpha_m )}{2  \alpha_m \cosh^2 \alpha_m \pi}
+  \frac{2  \cos^2 \alpha_m \pi  \tanh \alpha_m \pi  + \sin 2 \alpha_m \pi }{\alpha_m}. 
\end{equation}
We then define the normalized symmetric modes $\xi_m$, $m \geq 2$, even, by 
\begin{equation}
\label{norm-xi}
\xi_m(s) = \frac{1}{\tilde c_m}  {\tilde \xi}_m(s). 
\end{equation}
The values of the ${\cal C}_S(m)$, see  
\eqref{SymModes}, 
are  then $ {\cal C}_S(m) = ({\tilde c}_m \cosh \alpha_m \pi)^{-1}$, $m \geq 2$, even. 

Expression \eqref{normalization-form} indicates a feature seen also in the triple integrals, see Supplement, namely 
the integral has three types of terms: terms of $O(1)$ for large $m$, such as the factor $\pi$ in 
 \eqref{normalization-form},  terms of $O(\alpha^{-1}_m) \sim O(m^{-1})$ for $m$ large, 
 and terms that decrease exponentially in $m$  for $m$ large. 
The $O(1)$ contribution $\pi$ comes from the oscillatory part $\cos \alpha_m \pi$. This suggests 
approximating the $\xi_m$ in the triple integrals by their oscillating part, at least for   
high frequency mode interactions. 

The computation of the triple integrals
$I_{n_1'',n_2,n_3}$, $I_{n'_1, n'_2,n_3} $, $I_{n_1, n_2,n_3}$ of 
\eqref{spectral-coef-1},  \eqref{spectral-coef-2}
involves the normalized $\xi_m$, $m \geq 2$, and their first and second derivatives.   
Products of these functions can be integrated in closed form, 
see Supplement for the results. The final expressions are lengthy and must also be evaluated numerically. 
While we do see some patterns and possible simplifications, it appears practical 
to also  
evaluate the triple integrals numerically. To do this we use 
Clenshaw-Curtis quadrature where an integrand is approximated globally by a single polynomial 
that is then integrated exactly \cite{CC60}. 
The polynomial is obtained by interpolating the integrand at the Chebyshev-Lobatto nodes, see Supplement.  
The theory is further described in \cite{T00, E65, HS68, E93} and we indicate the accuracy of the method in the Supplement. 
For the modes we considered we 
typically obtain accuracy to $5-6$ significant digits using $N = 2^{10}$ nodes. 
The results of these calculations are shown in   
Tables \ref{tab:AandB}, \ref{tab:B} for the coefficients $\cal A$, $\cal B$ of  
\eqref{cubic-A-B-coeff} respectively. We have used $B = h = \pi$, $g = 1$. 
The first column of each table uses analytical expressions for the integrals 
$I_{n_1'',n_2,n_3}$, $I_{n'_1, n'_2,n_3} $, $I_{n_1, n_2,n_3}$; in the second column the integrals 
are evaluated using Clenshaw-Curtis quadrature, see Supplement. 

The values $B = h = \pi$, $g = 1$ were chosen for simplicity in \eqref{b-2-eq}-\eqref{a-4-eq}, 
and lead to results that are close to the physical scales for a domain of size $B = h = \pi$ meters.
We can check that results for $g \sim 9.8  m/sec^2$ in SI units can be obtained from  
the solutions of \eqref{b-2-eq}-\eqref{a-4-eq} below by 
dividing the time $t$ and $V$ by $\sqrt{9.8}$, i.e. the value of $g$. 
The results of the Supplement can be easily generalized for 
arbitrary values of $B$, $h$. Alternatively, observe that 
by \eqref{def-GE} the physical dimensions of $V$ are $\hbox{length}^2$, 
this fact can be used to derive adimensional forms of the equations.

\subsection{ \it Nonlinear mode interactions and spatial wave patterns}

In what follows we explore numerically some of the dynamics of the two-mode truncation \eqref{b-2-eq}-\eqref{a-4-eq}
and corresponding spatial shapes of the surface. 
%
We use 
a fourth-order Runge-Kutta scheme with a numerical time step $\Delta t= 10^{-6}$. 
The method is fourth-order accurate in time, giving a $O(\Delta t^4)$ local error for the full time evolution.
The numerical scheme was implemented using Fortran. 
 
\begin{figure}
    \centering
      \includegraphics[scale=0.34]{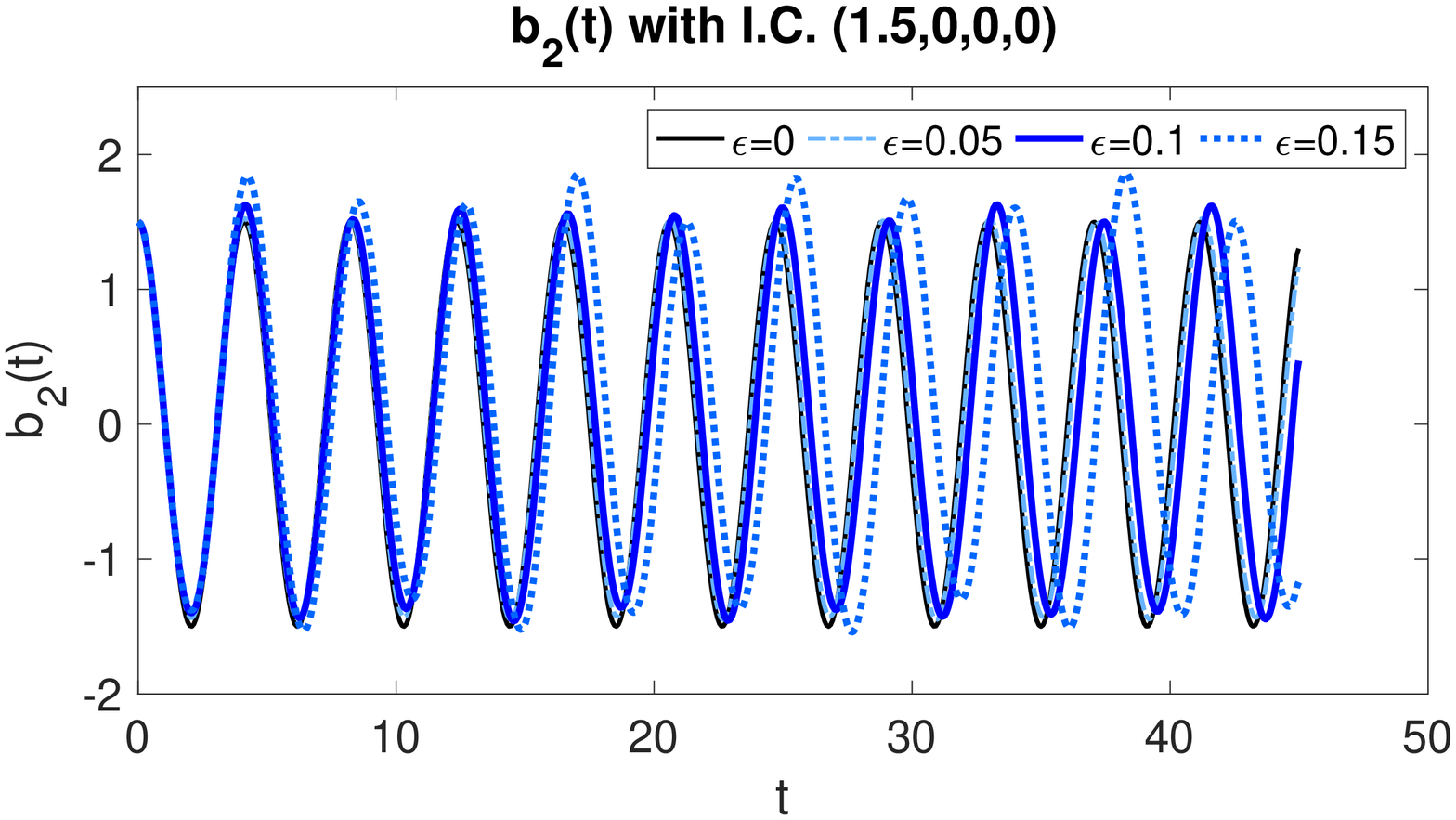}
       \includegraphics[scale=0.34]{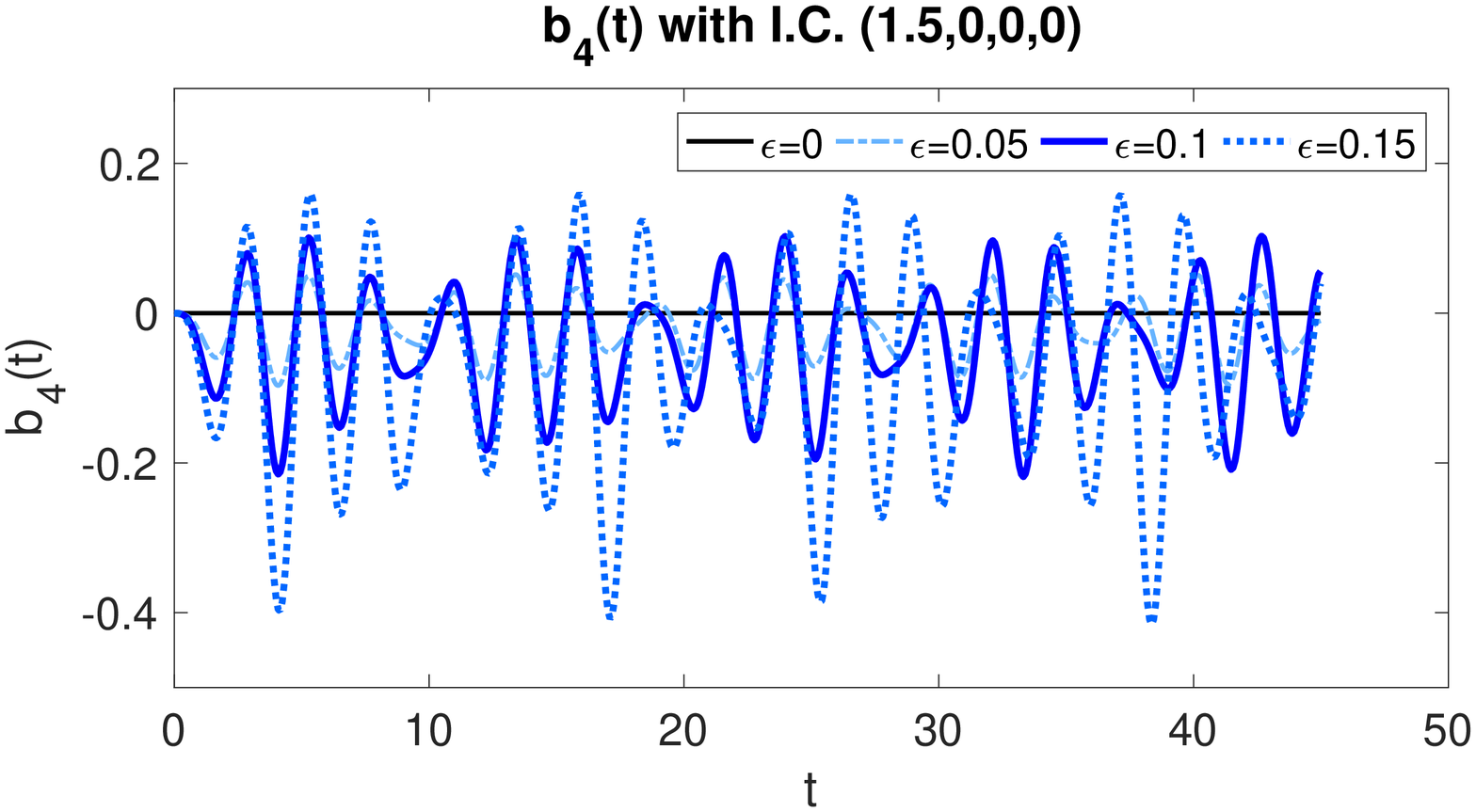}
       \includegraphics[scale=0.34]{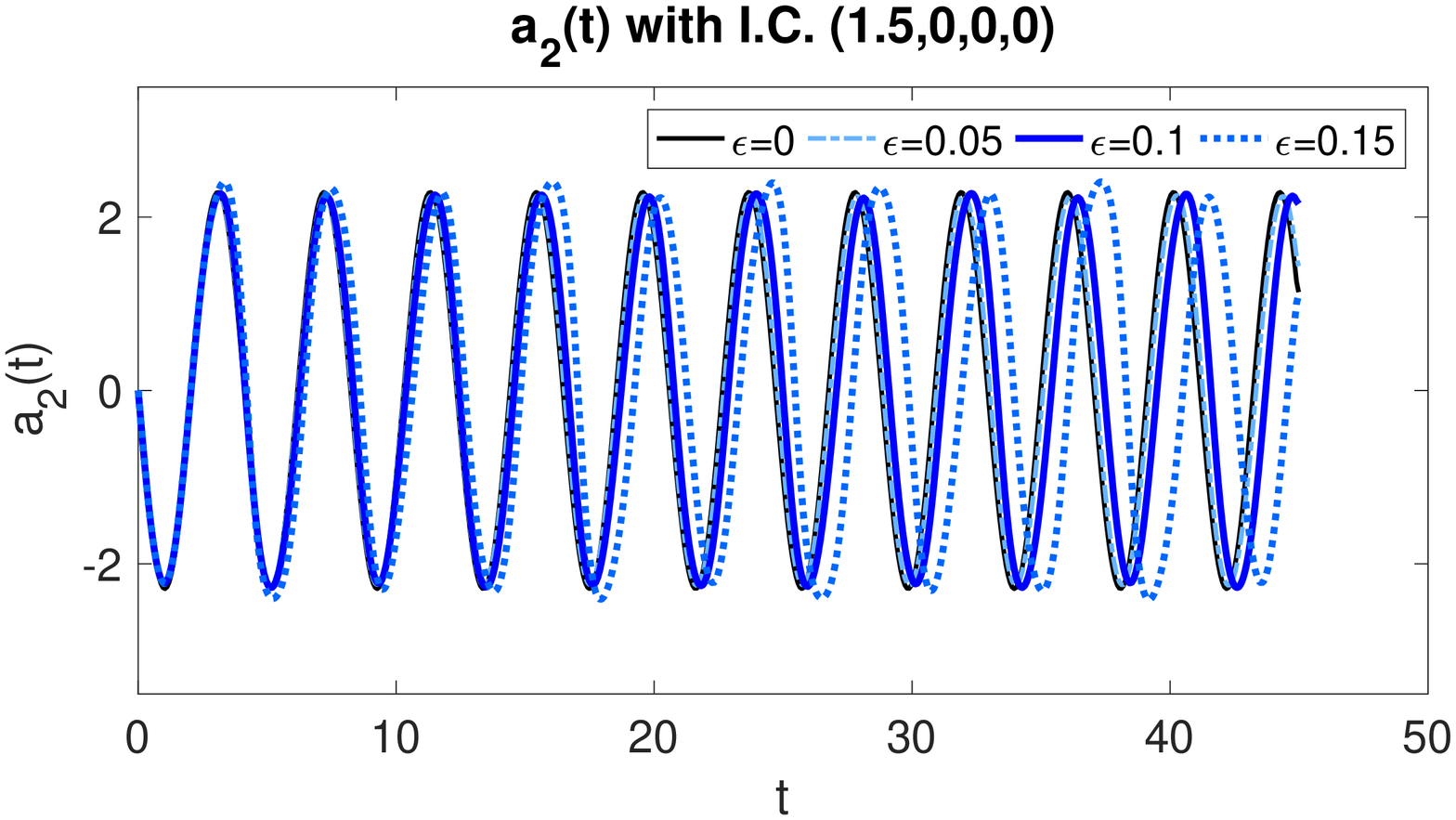}
        \includegraphics[scale=0.34]{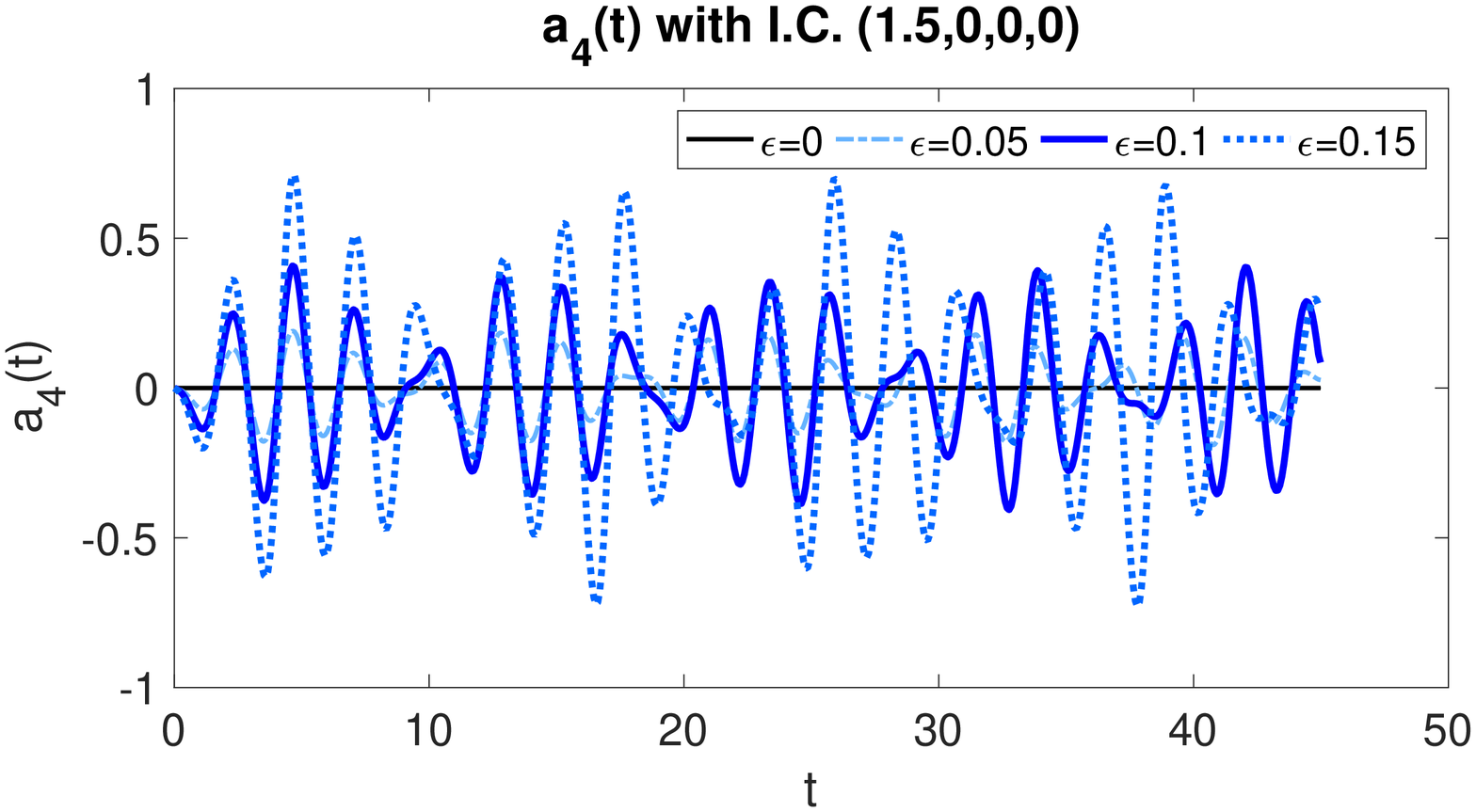}
    \caption{Solutions $b_2(t)$, $b_4(t)$, $a_2(t)$, $a_4(t)$ 
    of \eqref{b-2-eq}-\eqref{a-4-eq}, $t\in [0,45]$, 
    for initial condition $(b_1,b_3,a_1,a_3)= (1.5,0,0,0)$. Black line: $\epsilon=0$. Light blue dashed line: $\epsilon=0.05$. Blue line: $\epsilon=0.1$. Gray blue dotted  line: $\epsilon=0.15$. }
    \label{fig:b2-b4-a2-a4-index2-1p5}
\end{figure}

\begin{figure}
    \centering
            \includegraphics[scale=0.34]{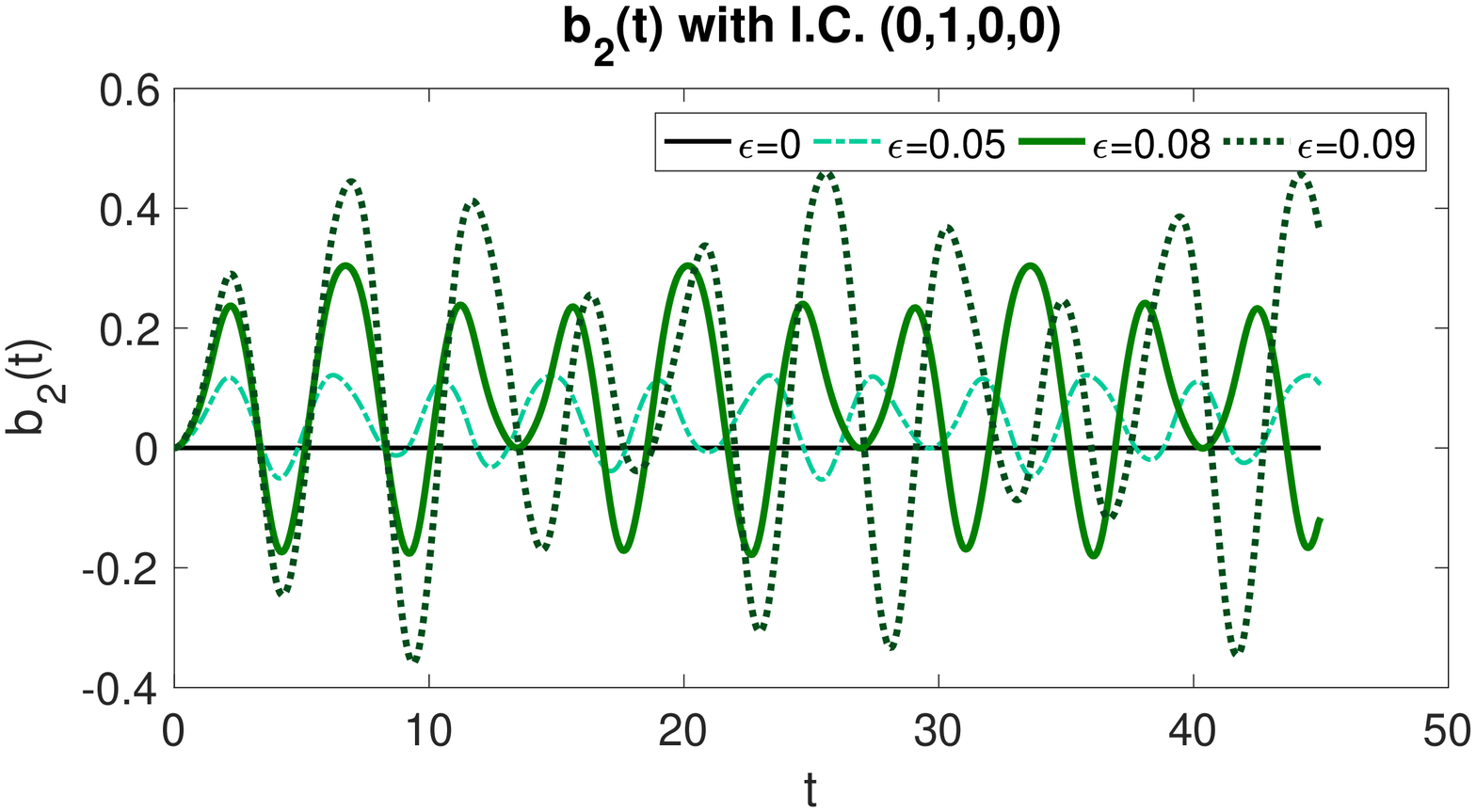}
      \includegraphics[scale=0.34]{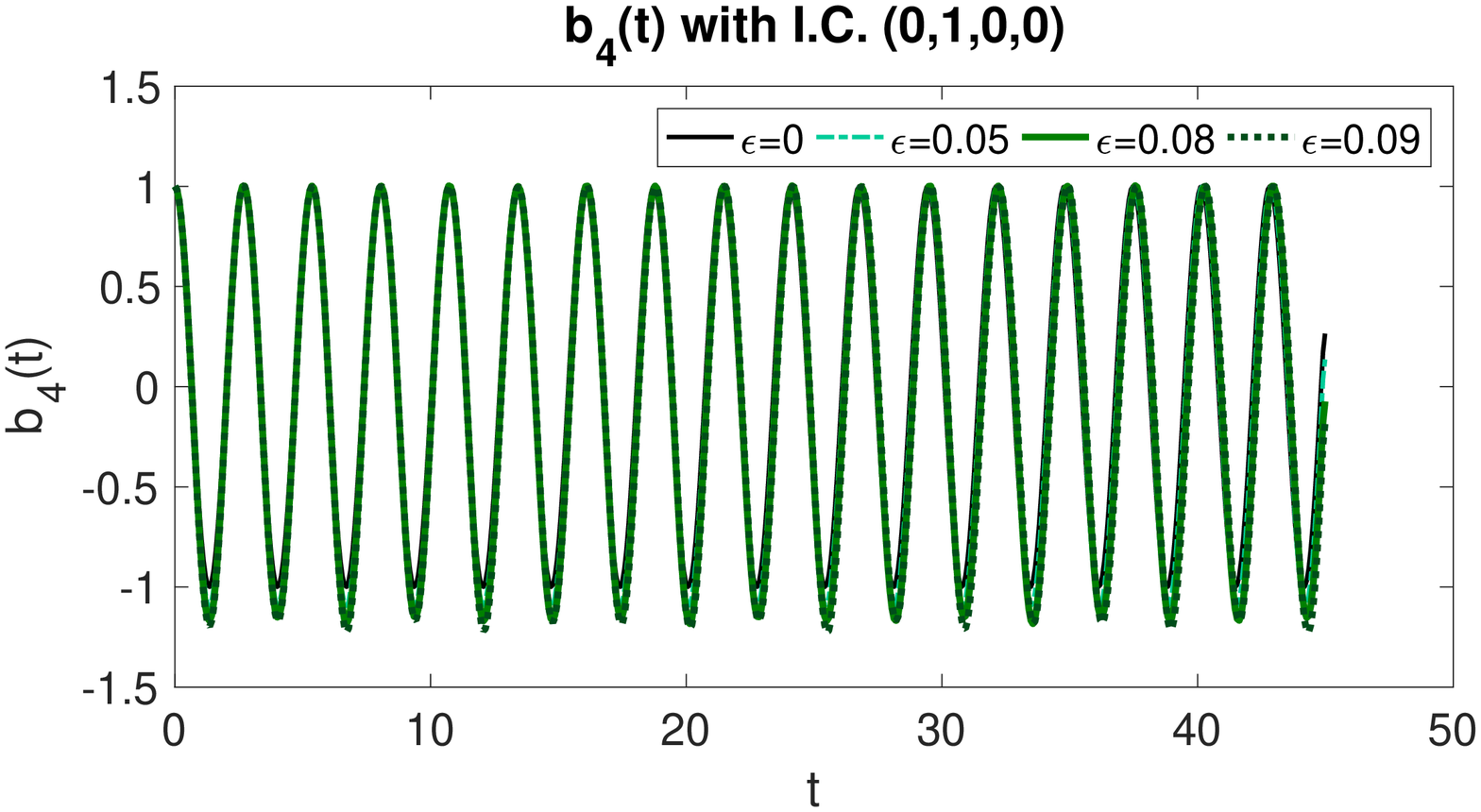}
      \includegraphics[scale=0.34]{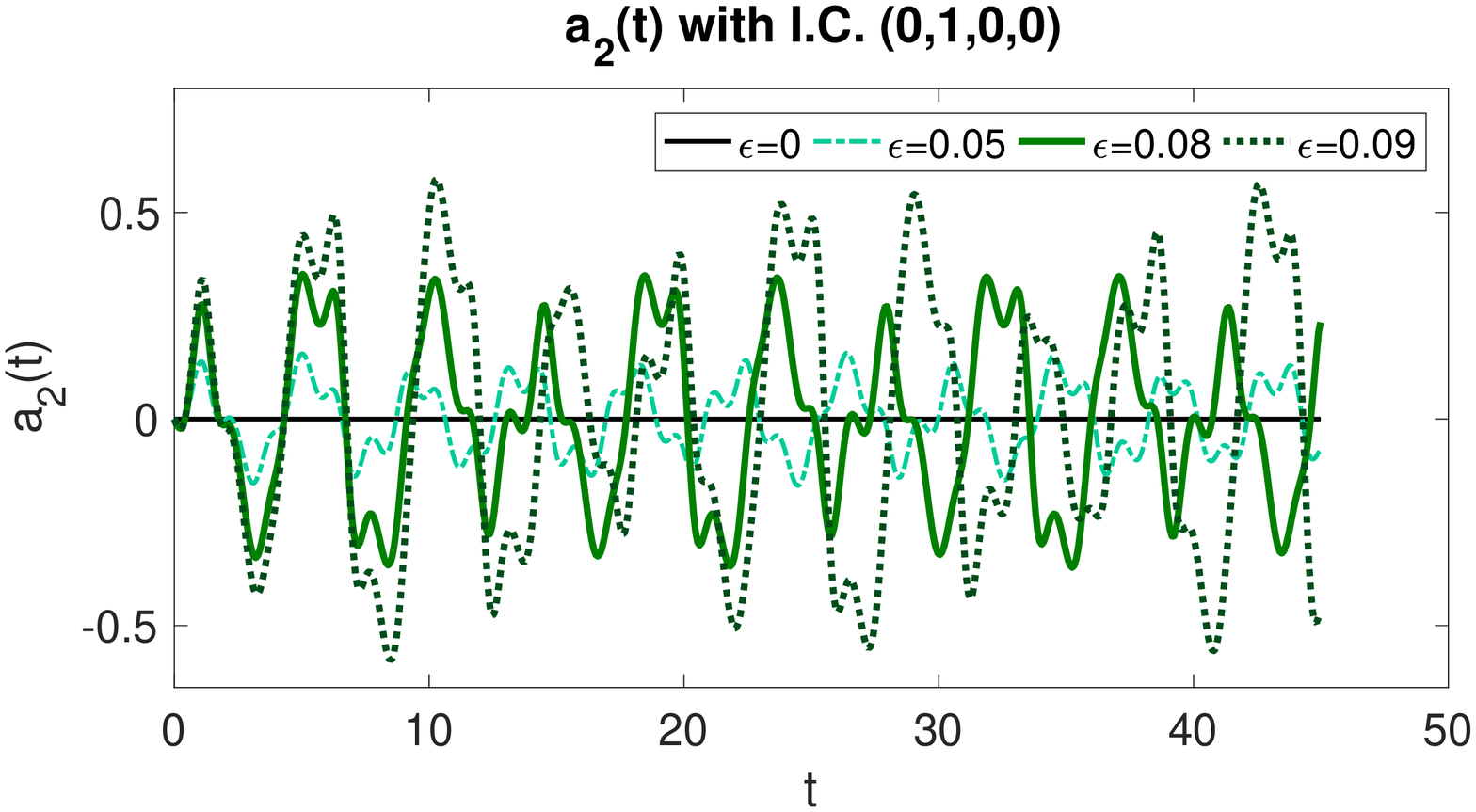}
       \includegraphics[scale=0.34]{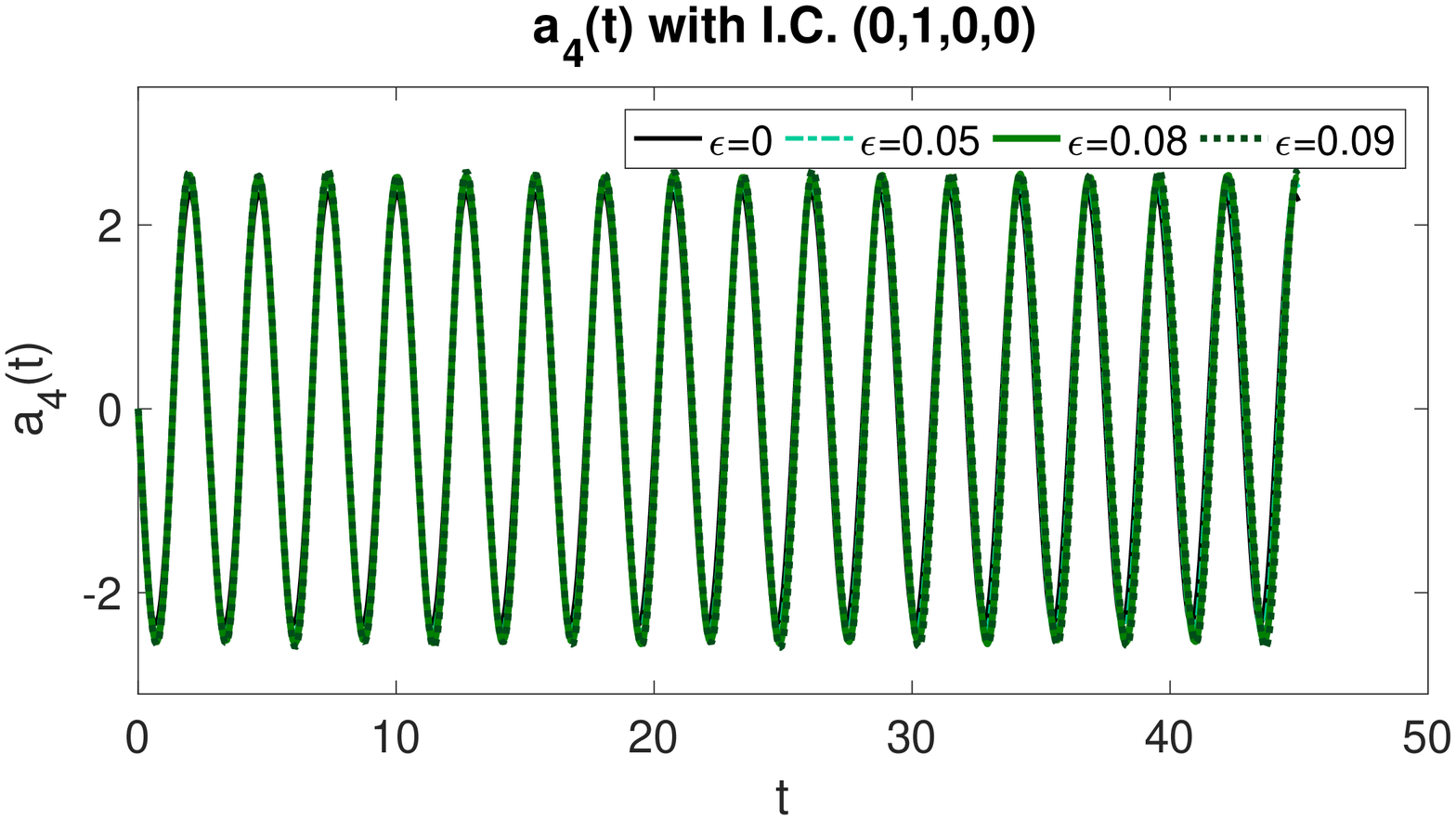}      
    \caption{Solutions $b_2(t)$, $b_4(t)$, $a_2(t)$, $a_4(t)$
 of \eqref{b-2-eq}-\eqref{a-4-eq}, $t\in [0,45]$,     
    for initial condition $(b_1,b_3,a_1,a_3)= (0,1,0,0)$. Black line: $\epsilon=0$. Light green dashed line: $\epsilon=0.05$. Green line: $\epsilon=0.08$.  Dark green dotted line: $\epsilon=0.09$.}
    \label{fig:b2-b4-a2-a4-index4-1}
\end{figure}

Equations \eqref{first-spectral-eq}, \eqref{second-spectral-eq}
and their truncations do not have an apparent Hamiltonian structure and a basic question 
is the long-time existence of solutions of these truncations. 
We consider this question examining first near-monochromatic initial conditions. 
We see that trajectories can exist for several multiples of the period of the slowest mode without apparent 
increase in amplitude. In the figures we show the evolution 
for $ t \in [0,45]$ ($ \sim 10 $ periods of the $n = 2$ mode).
Figure \ref{fig:b2-b4-a2-a4-index2-1p5} 
shows solutions
where most of the amplitude is in mode $n =2 $ (the initial amplitude in mode $n =4$ vanishes), while 
Figure \ref{fig:b2-b4-a2-a4-index4-1} shows
solutions where most of the amplitude is in mode $n =4$ (the initial amplitude in mode $n =4$ vanishes).
In both cases we examine different initial amplitudes, and we also vary $\epsilon$ (this variation can be also 
absorbed in the initial amplitude). 
We see that the evolution of the dominant mode is nearly sinusoidal, with a weak modulation as we increase the initial amplitude. 
The motion of the other mode has a more complicated shape by its amplitude remains relatively small. The evolution of the 
smaller mode can be modeled by system \eqref{one-mode-b-eq}-\eqref{one-mode-a-eq} with forcing terms that are approximately periodic.

\begin{figure}
    \centering
      \includegraphics[scale=0.34]{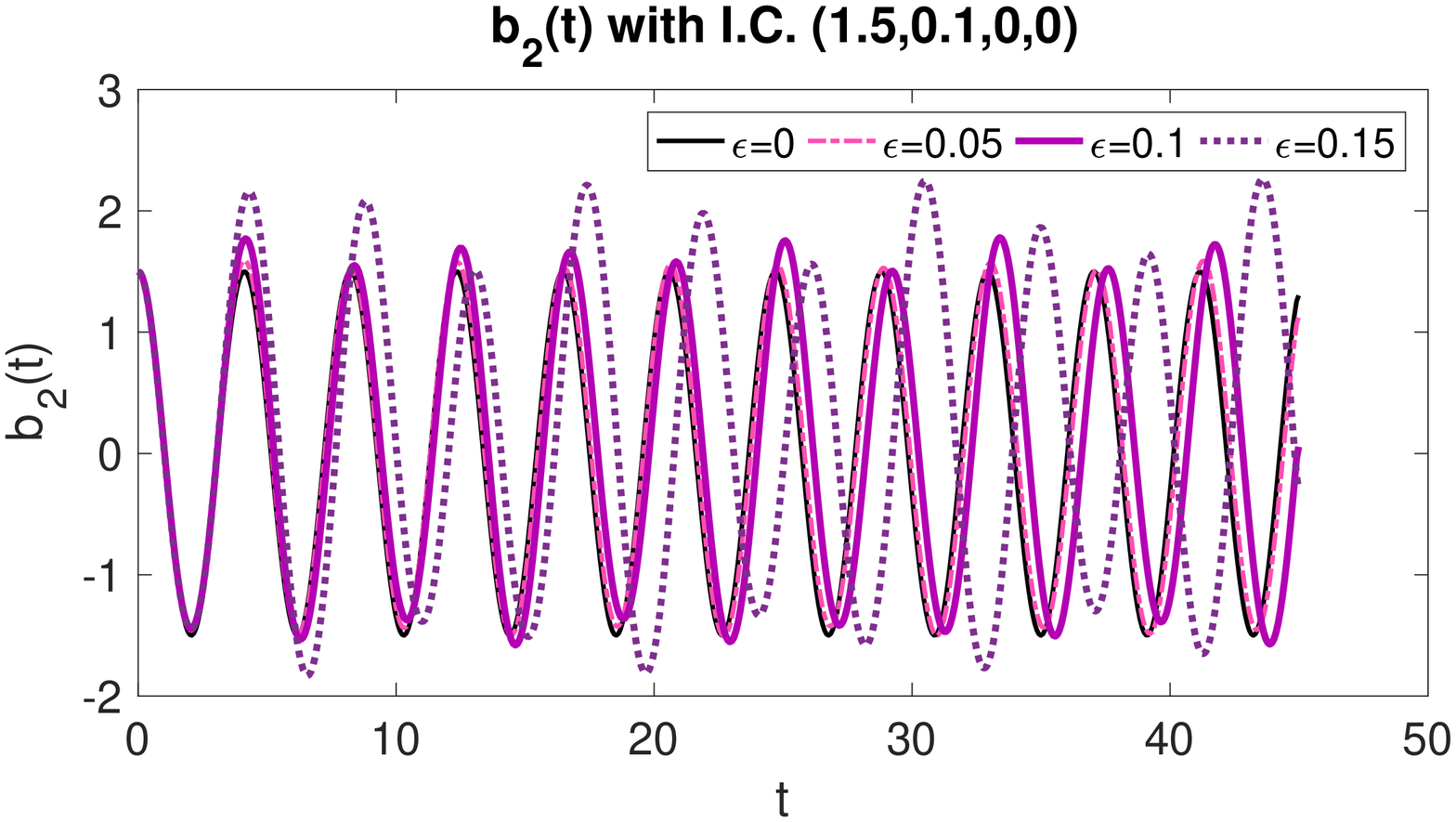}
       \includegraphics[scale=0.34]{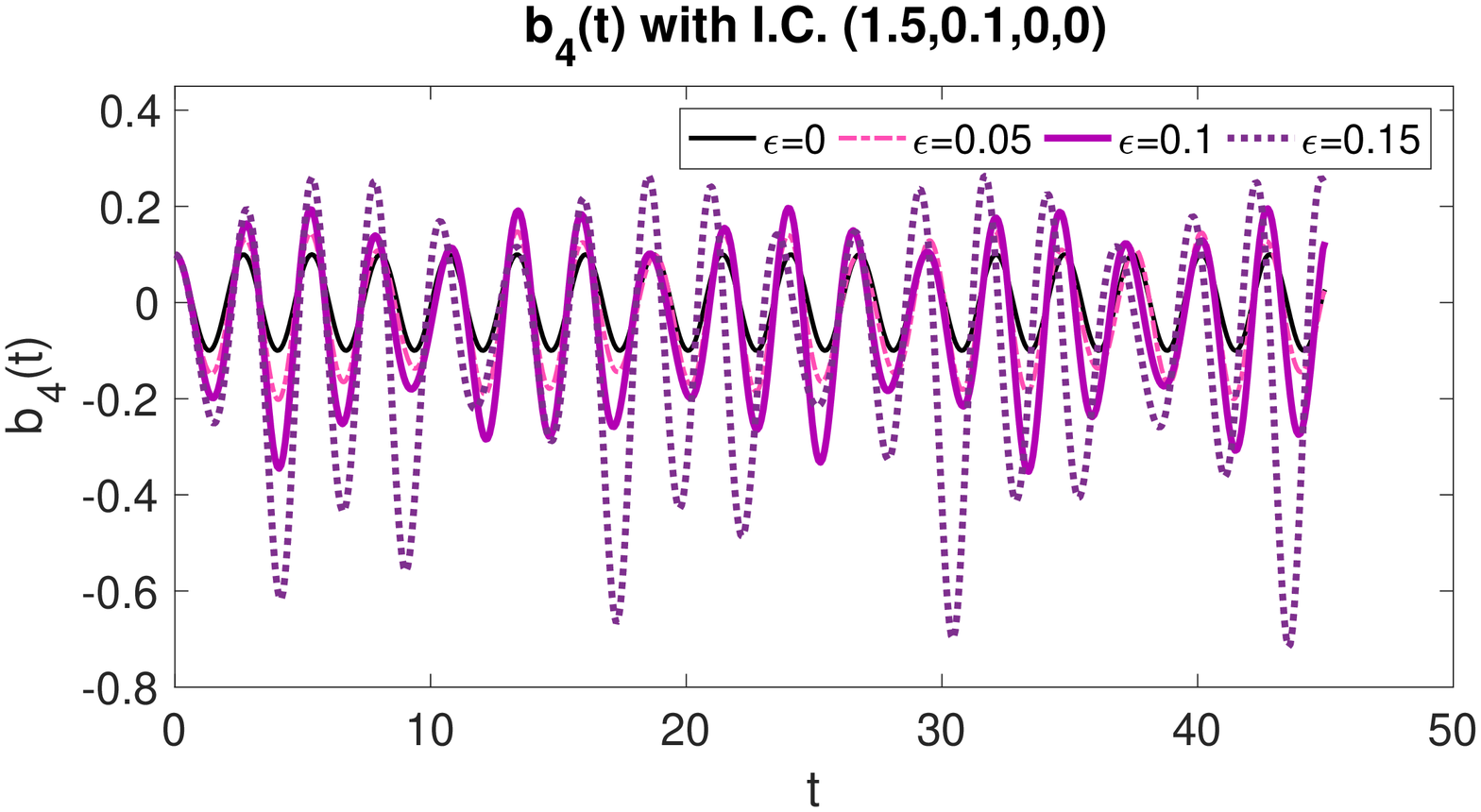}
       \includegraphics[scale=0.34]{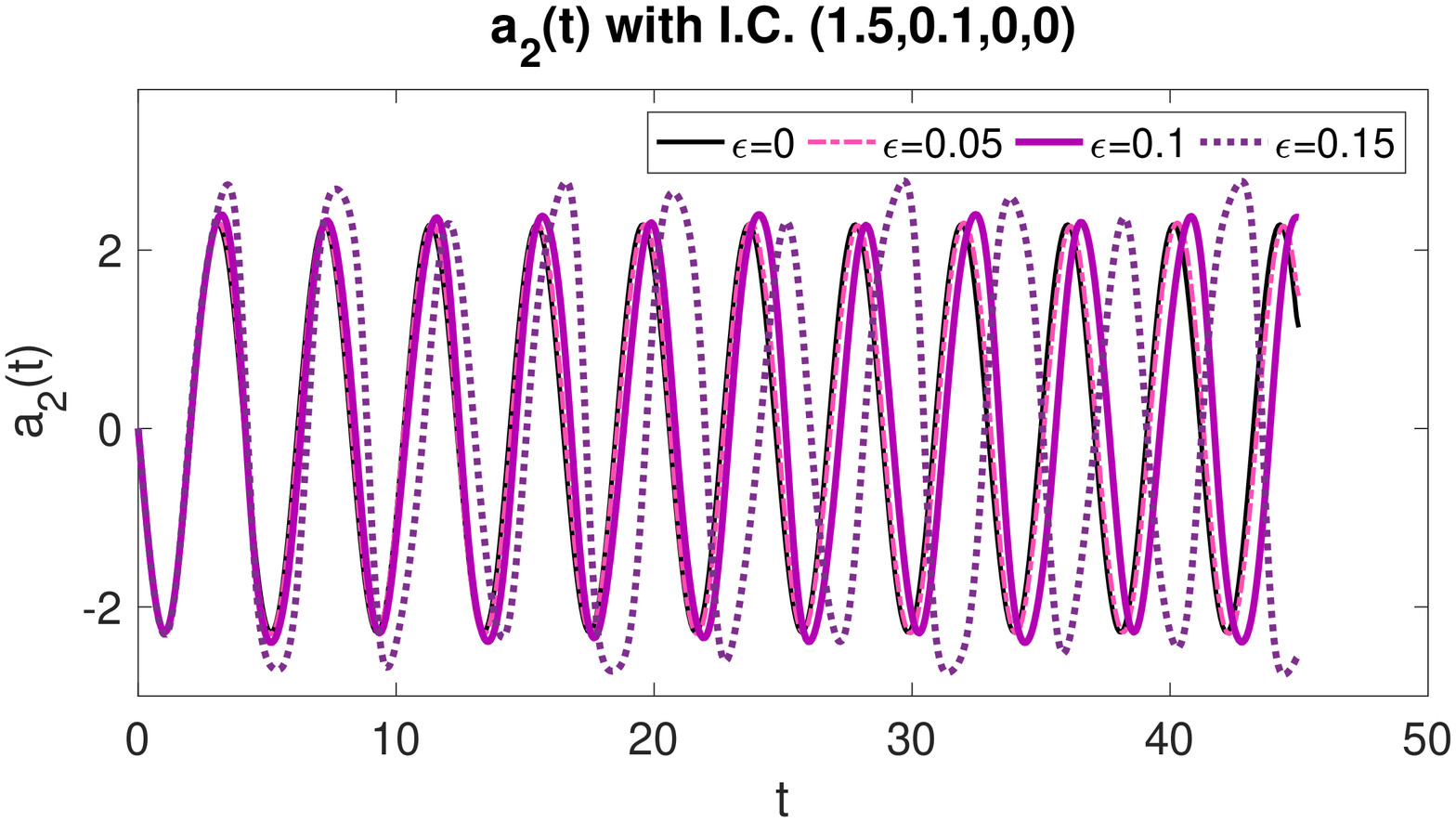}
       \includegraphics[scale=0.34]{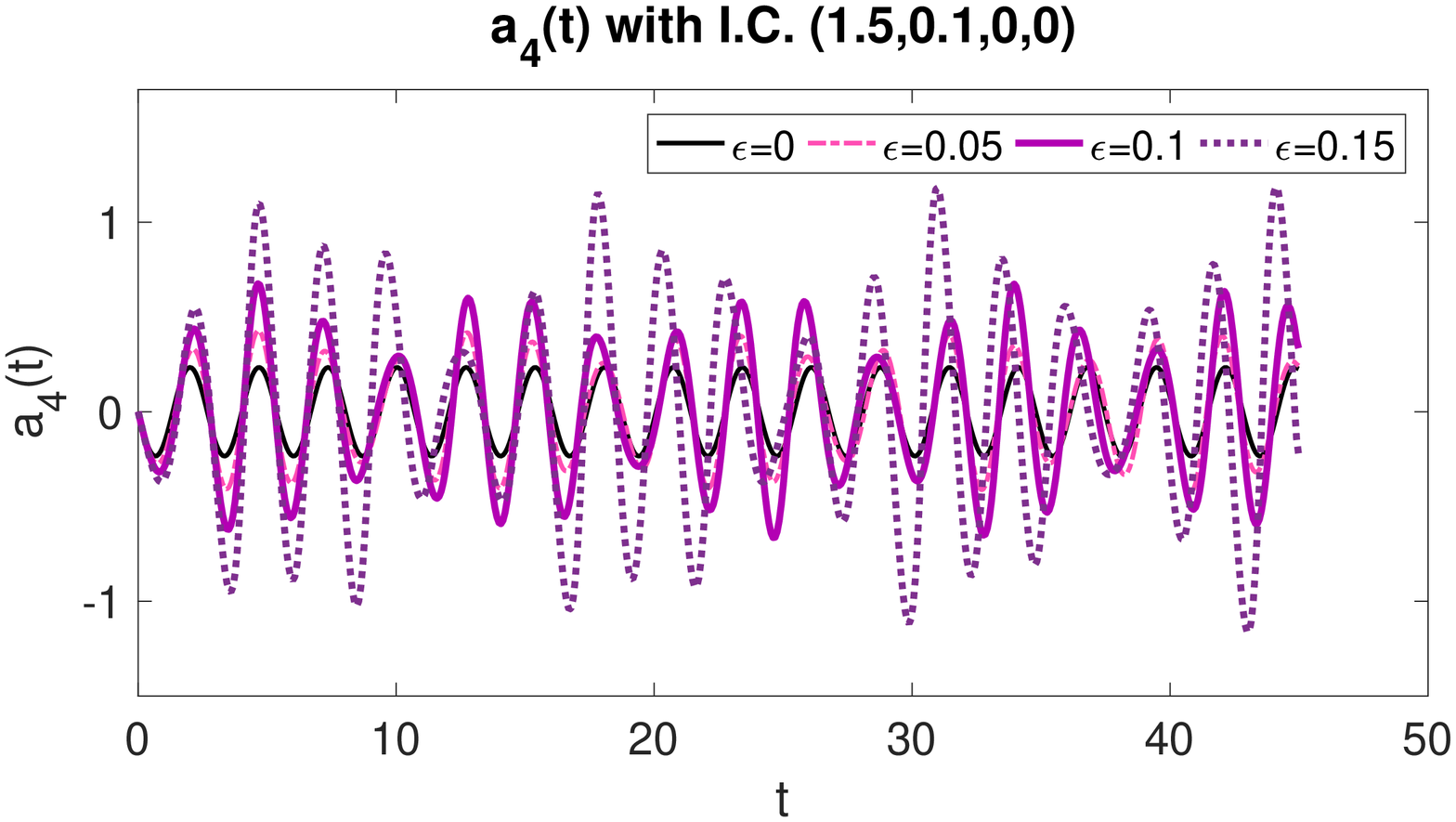}
    \caption{Solutions $b_2(t)$, $b_4(t)$, $a_2(t)$, $a_4(t)$
of \eqref{b-2-eq}-\eqref{a-4-eq}, $t\in [0,45]$,          
    for initial condition $(b_1,b_3,a_1,a_3)= (1.5,0.1,0,0)$. Black line: $\epsilon=0$. Pink dashed line: $\epsilon=0.05$. Violet line: $\epsilon=0.1$. Deep purple dotted  line: $\epsilon=0.15$.}
    \label{fig:b2-b4-a2-a4-index2and4-1p5}
\end{figure}

Figure 
\ref{fig:b2-b4-a2-a4-index2and4-1p5}
shows 
the evolution from initial conditions where the amplitude of modes $n =2$, and $4$ are closer. 
At small amplitudes we see negligible interaction. 
Increasing the amplitude we 
see considerable modulation of both modes.
 
The nonlinear effect of mode interactions is therefore seen clearly for higher amplitudes. At the same time these nonlinear effects  
are seen in solutions that exist over several periods of the linear motion. A possible theoretical explanation of boundedness 
of small amplitude solutions is outside the scope of this work but is a problem we may consider in the future. 

The range of initial amplitudes and $\epsilon$ in the figures above is chosen so as to avoid 
rapidly growing solutions. Moreover, numerical integrations were performed for times up to at least $t = 135$ for all 
initial conditions of the figures above. In all cases 
we see qualitatively similar amplitude modulation without any indication of larger amplitudes or unbounded growth.   
Examples of trajectories that seem to grow without bound are also seen but were not studied in detail. 
We note that the one-mode system 
\eqref{one-mode-b-eq}-\eqref{one-mode-a-eq} has nontrivial fixed points of size $O(\epsilon^{-1})$ 
that could lead to unbounded trajectories.  
More detailed dynamical studies of such phenomena in truncations can be pursued further. Unbounded orbits are likely  
an artifact of the quadratic nonlinearity, but are still of interest since they indicate the limitations of the quadratic model. 
Higher order equations are also of interest but are more cumbersome.

The numerically computed mode amplitudes $b_2(t)$, $b_4(t)$, $a_2(t)$, $a_4(t)$
are used to obtain the free surface using the spectral truncation $ {\cal G}^{\epsilon}_{E,T} (s,t) $
of the approximate expression \eqref{GE-surface-approx}, namely  
  \begin{equation}
\label{trunctated-GEtilde}
    {\cal G}^{{\epsilon}}_{E,T} (s,t) =  
    [s +{\epsilon} \sum_{n \in \{2,4\}} b_n(t) \partial_x \Phi_n([s,\pi]),
    \pi +  {\epsilon} \sum_{n \in \{2,4\}}b_n(t) \partial_y \Phi_n([s,\pi)], 
\end{equation}
$s \in [-\pi,\pi] $. 
Note that the parameter $\epsilon$ appearing in  expressions for the surface, e.g. 
\eqref{GE-surface-approx}, and the spectral evolution equations \eqref{first-approx-syst},
\eqref{second-approx-syst} 
and their truncations is the same.  We may however compare the spatial shape of solutions of different dynamical 
equations by choosing different values for $\epsilon$ in the free surface and in the dynamical 
equations. 

\begin{figure}
      \centering
      \includegraphics[scale=0.25]{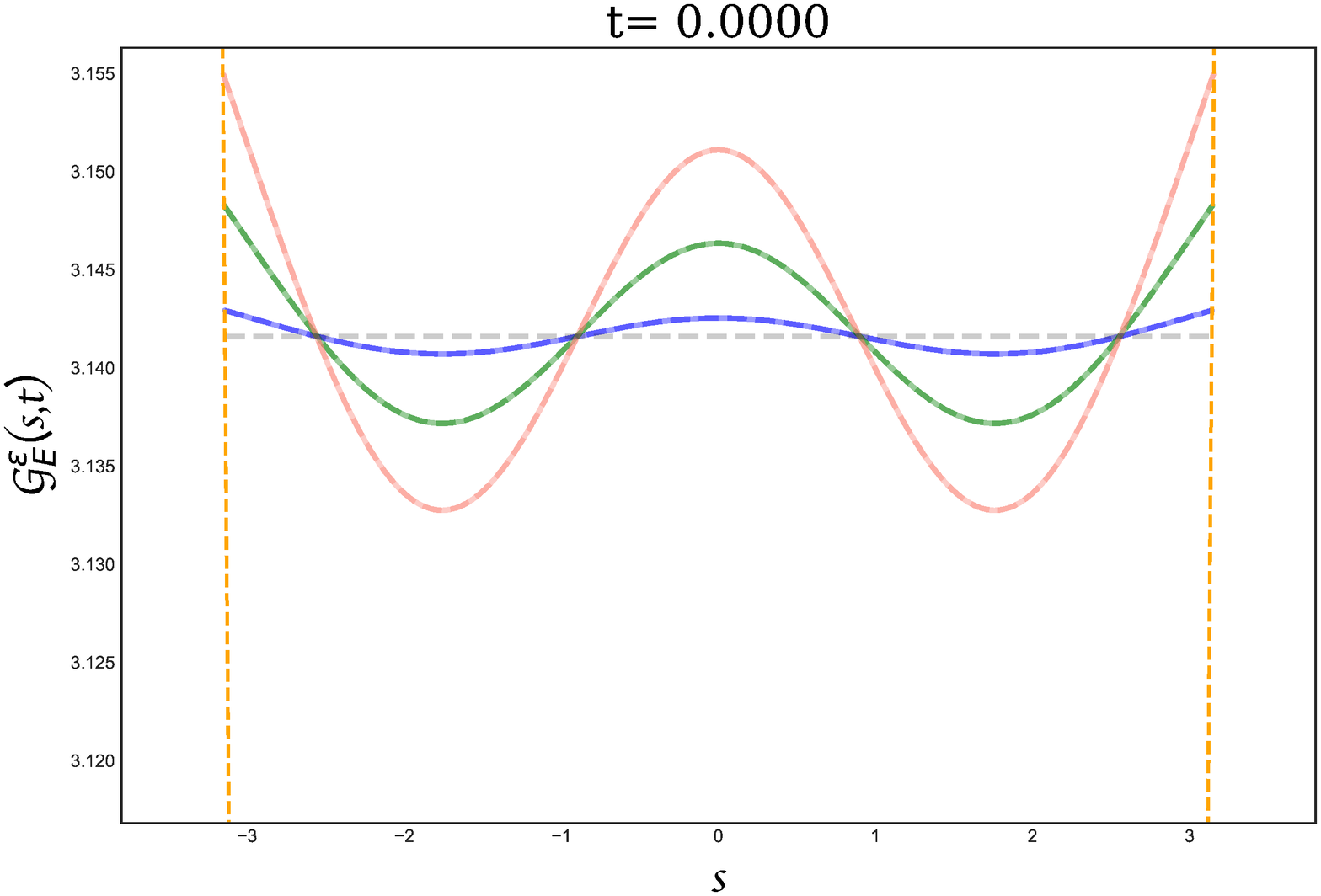}
        \includegraphics[scale=0.25]{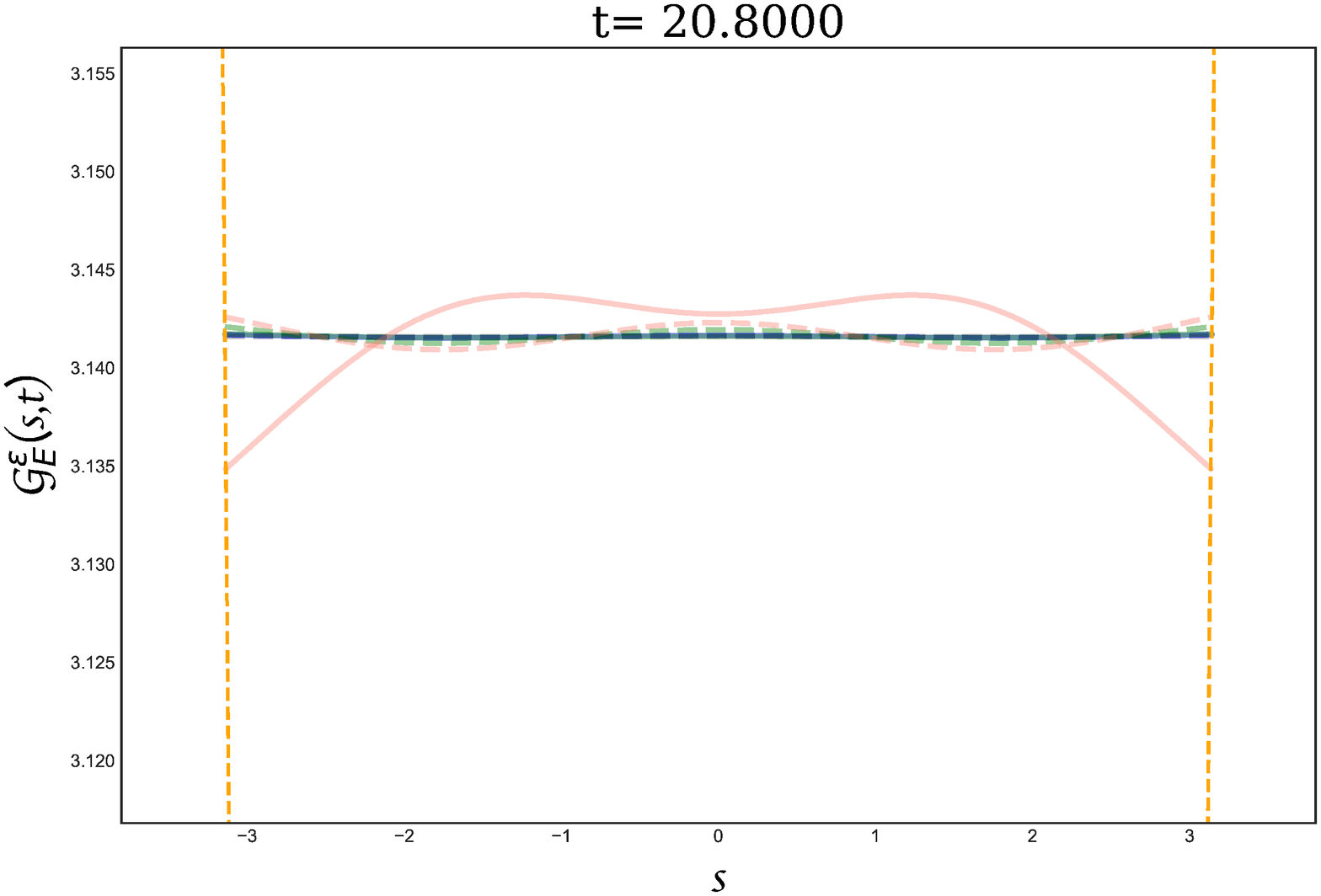}
       \includegraphics[scale=0.25]{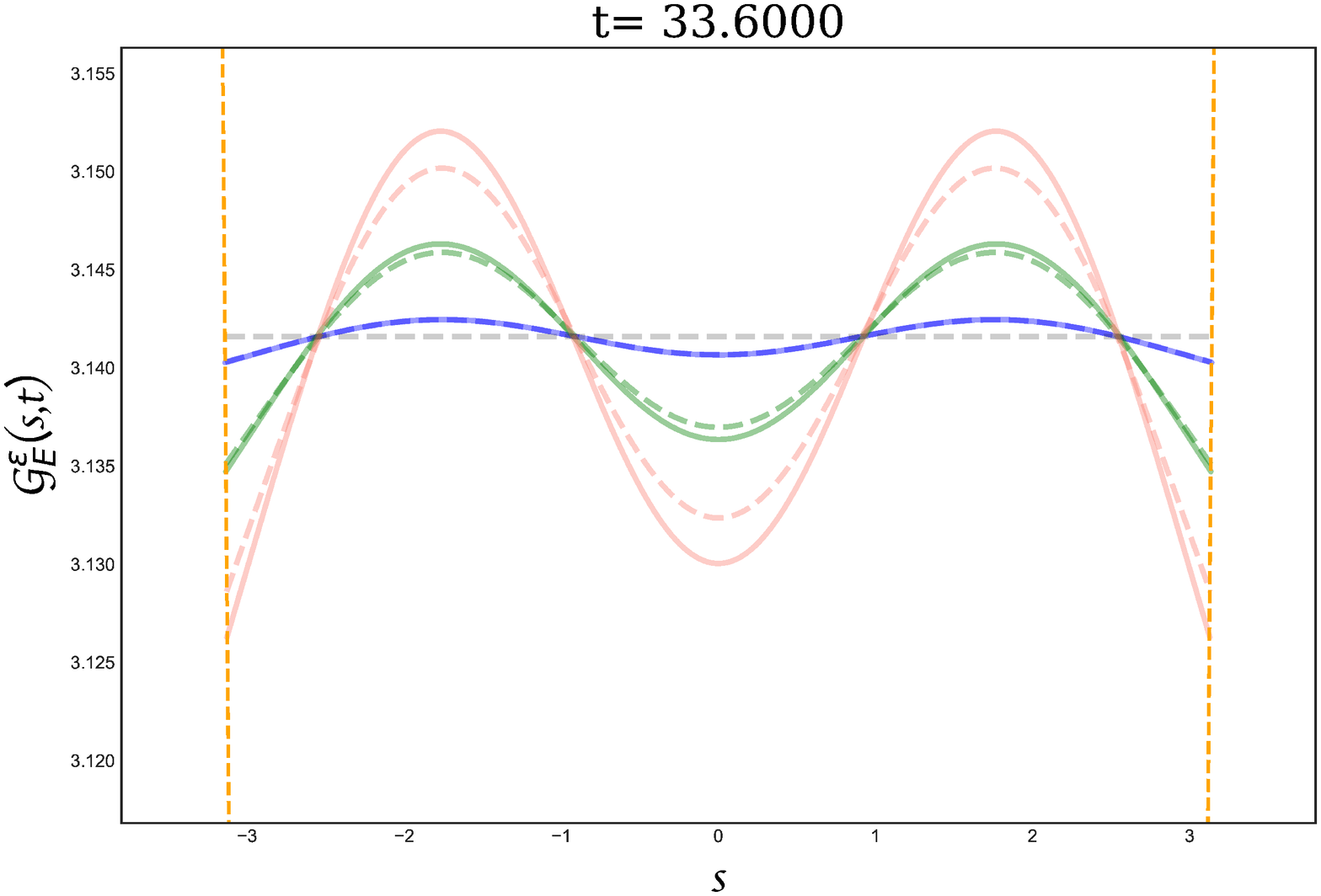}
        \includegraphics[scale=0.25]{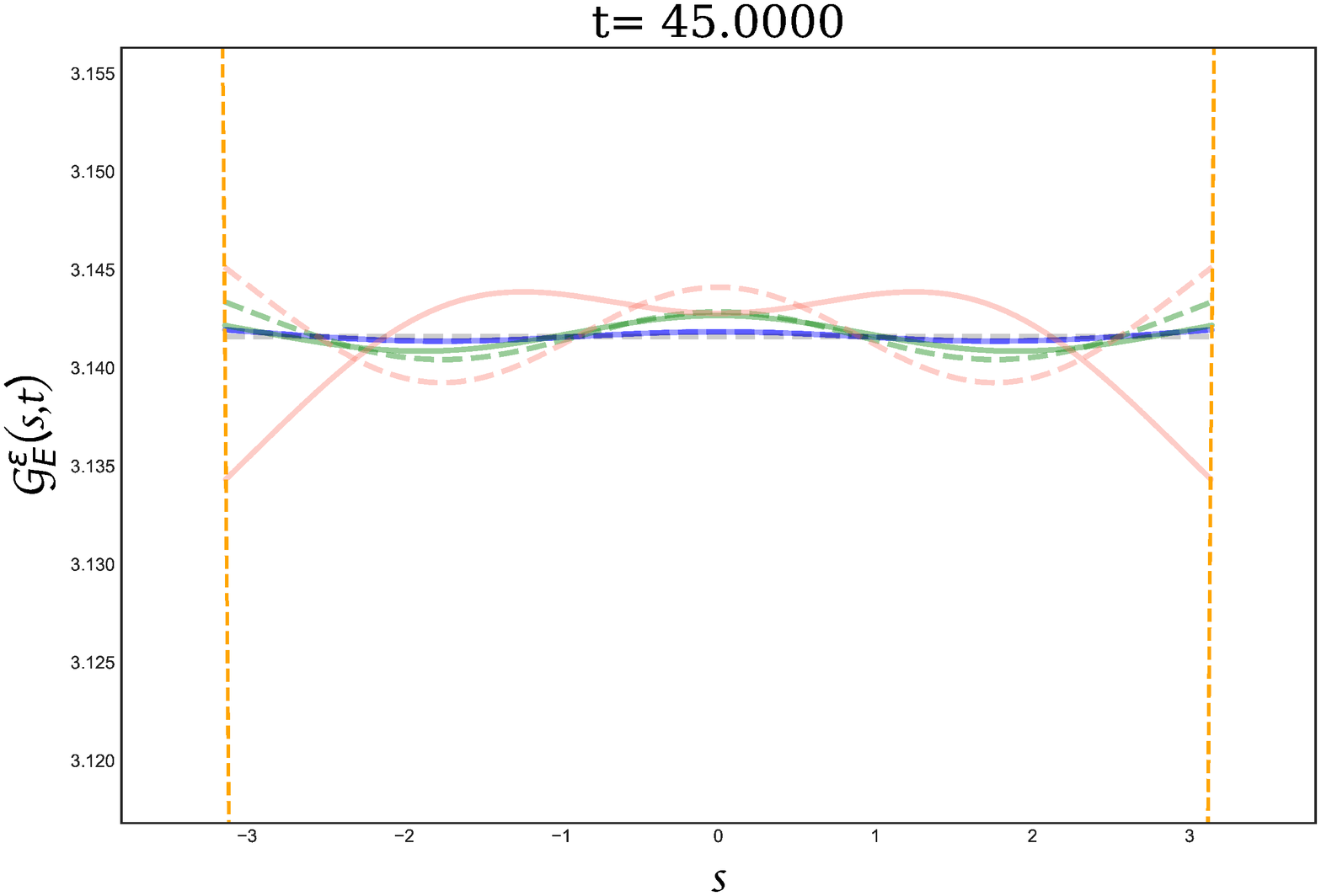}
    \caption{Surface profile of solutions at different times $t$.  
         Solid lines:      
    blue, $\epsilon = 0.09$, initial conditions $(0,0.1,0,0)$; green, $\epsilon = 0.09$, initial conditions 
    $(0,0.5,0,0)$; pink, $\epsilon = 0.09$, initial conditions $(0,1,0,0)$ (see Fig. 3).   
 Dashed lines 
         represent linear evolution for each initial condition.}   
    \label{fig:WallsCaso1b}
\end{figure}
\begin{figure}
      \centering
              \includegraphics[scale=0.4]{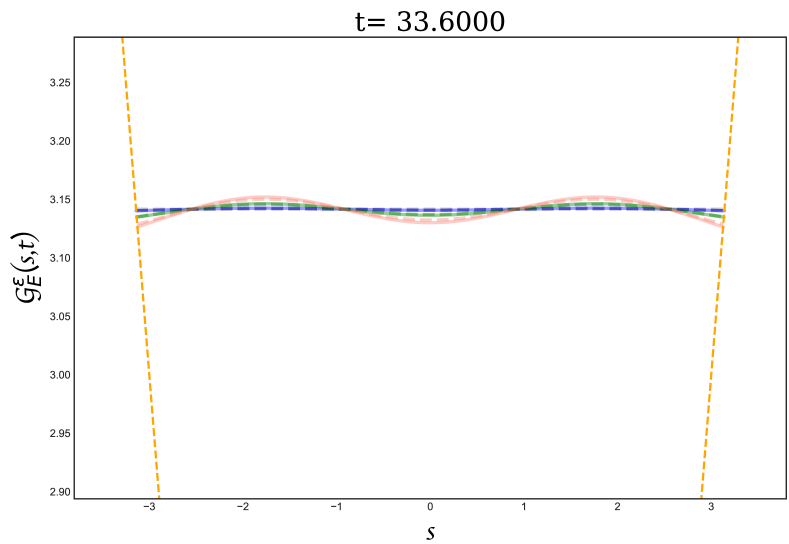}
       \caption{Zoom-out of Figure 14 at $t = 33.6 $.        
       }
    \label{WallsCaso1bZoomOut}
\end{figure}

Figures \ref{fig:WallsCaso1b}, \ref{WallsCaso1bZoomOut}
show snapshots of the free surface for the near-monochormatic waves where the mode $n = 4 $ 
dominates, see 
Figure \ref{fig:b2-b4-a2-a4-index4-1} with $\epsilon = 0.09$. 
The dashed lines show the surface for 
a linearized model, 
i.e. $\epsilon = 0.1$ in \eqref{trunctated-GEtilde}, and 
$\epsilon = 0 $ in the mode evolution equations \eqref{first-approx-syst}-\eqref{second-approx-syst}. 
Figure \ref{fig:WallsCaso1b} shows clearly how the boundary of the free surface follows 
the wall. The vertical scale used makes the wall appear almost vertical. 
In Figure 
\ref{WallsCaso1bZoomOut} we zoom out one these snapshots using a vertical scale that makes the inclination of the wall more obvious.

%
%

\begin{figure}
    \centering
       \includegraphics[scale=0.38]{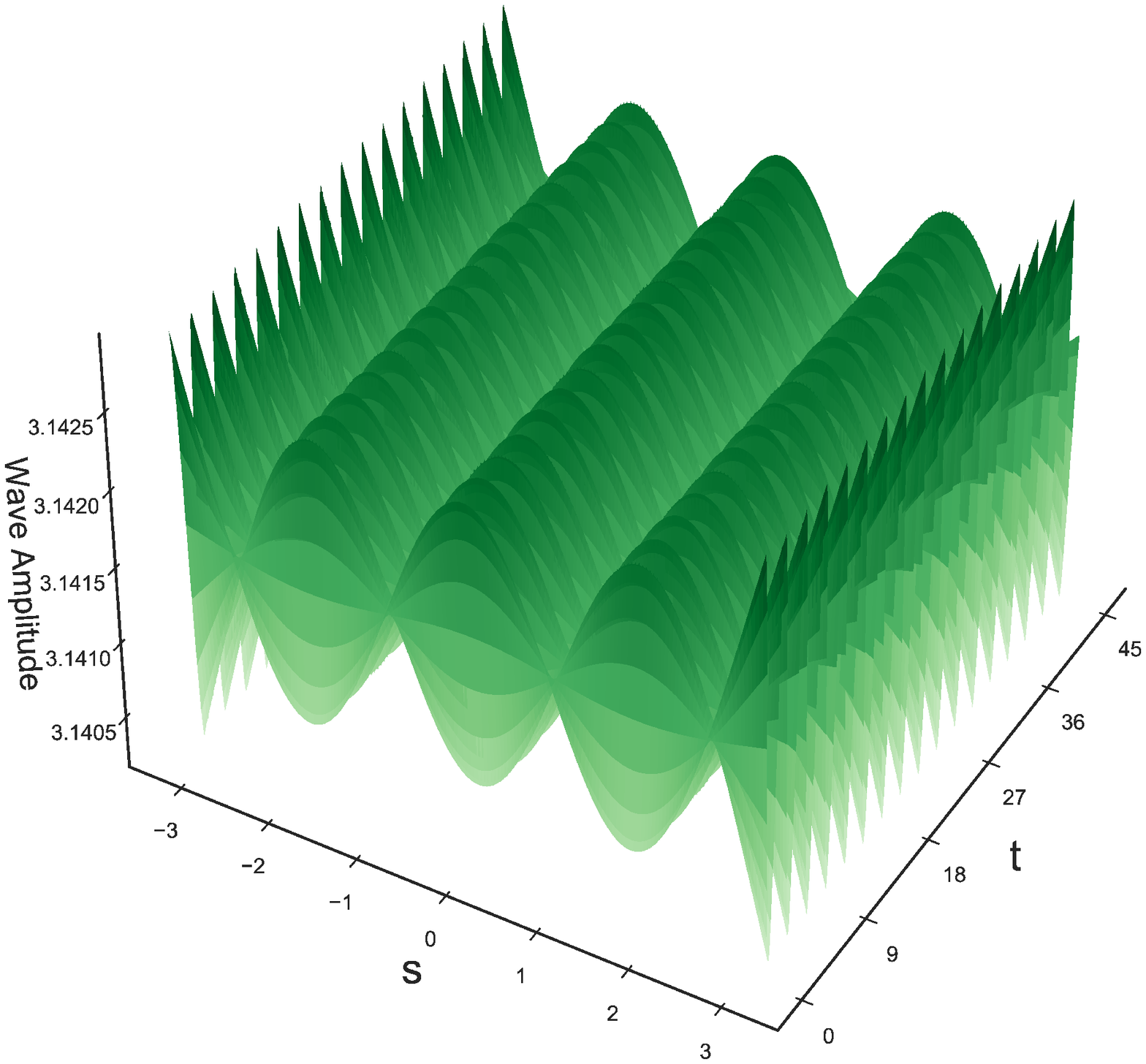}
       \includegraphics[scale=0.38]{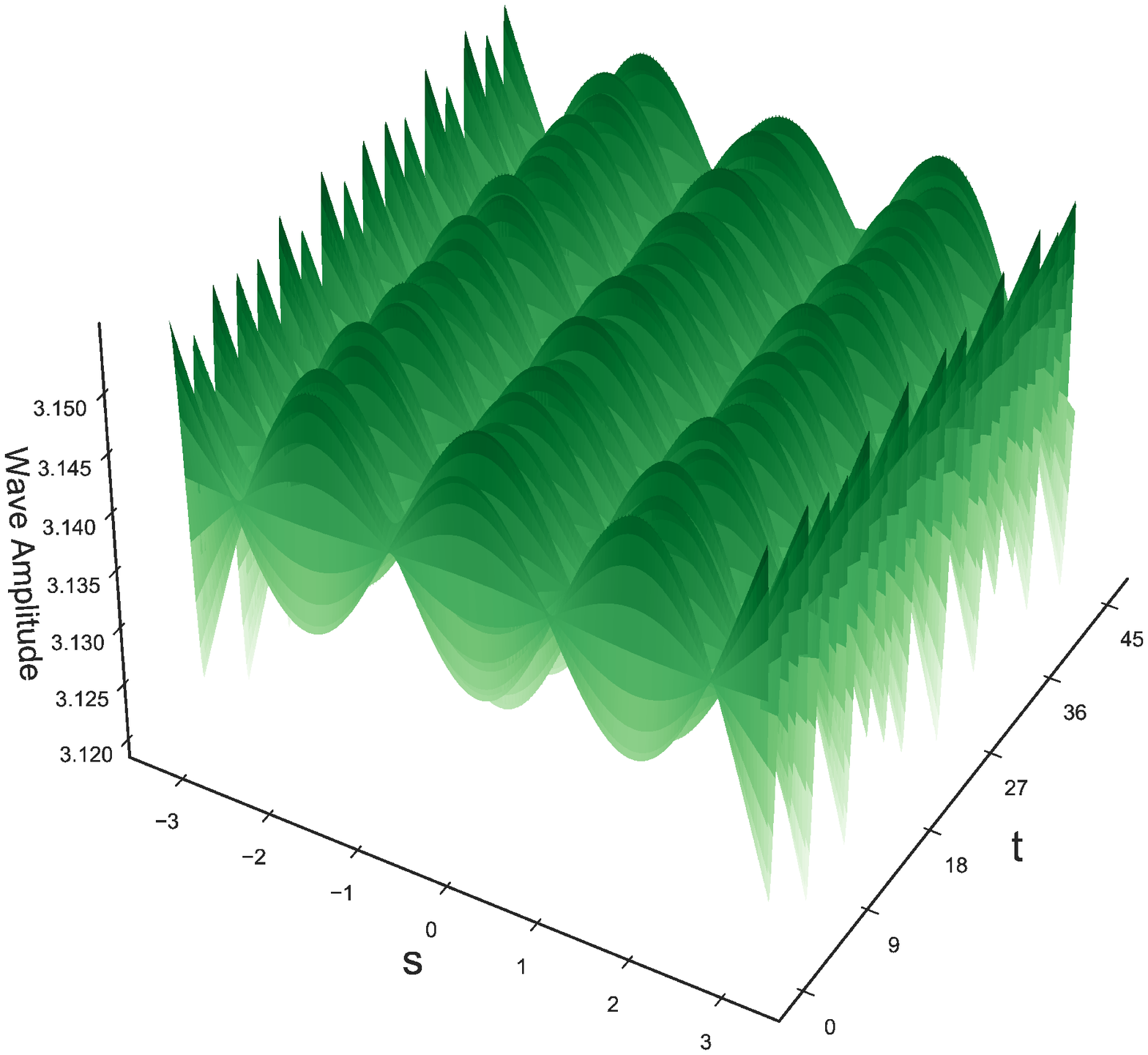}
    \caption{Spatiotemporal patterns of solutions: 
    \textbf{Left:} initial condition $(b_2,b_4,a_2,a_4)= (0,0.1,0,0)$, $\epsilon =0.09$. 
    \textbf{Right:} initial condition $(b_2,b_4,a_2,a_4)=(0,1,0,0)$, $\epsilon =0.09$ (see Fig. 3, dark green dotted line).}
    \label{fig:surfaceMode4}
\end{figure}

The expression for the free surface 
\eqref{trunctated-GEtilde} at different times allows us also to examine the spatiotemporal patterns
implied by the evolution of the mode amplitudes and the shape of the normal modes. 
Spatiotemporal patterns for the two near-monochromatic waves with dominant mode 
$4$ 
are shown in Figure 
\ref{fig:surfaceMode4}. 
Figure
\ref{fig:surfaceMode4} (a) shows the smaller amplitude motions.
The spatial pattern corresponds to harmonic oscillation of a single mode.  
In 
Figure 
\ref{fig:surfaceMode4} (b) we show the higher amplitude motions
that correspond to the mode evolution of 
Figure 
\ref{fig:b2-b4-a2-a4-index4-1}.
We see higher amplitude motions and 
a more complicated spatial pattern due to the presence of the second mode. 

\begin{figure}
      \centering
         \textbf{Wave run-up}\par\medskip

     \includegraphics[scale=0.31]{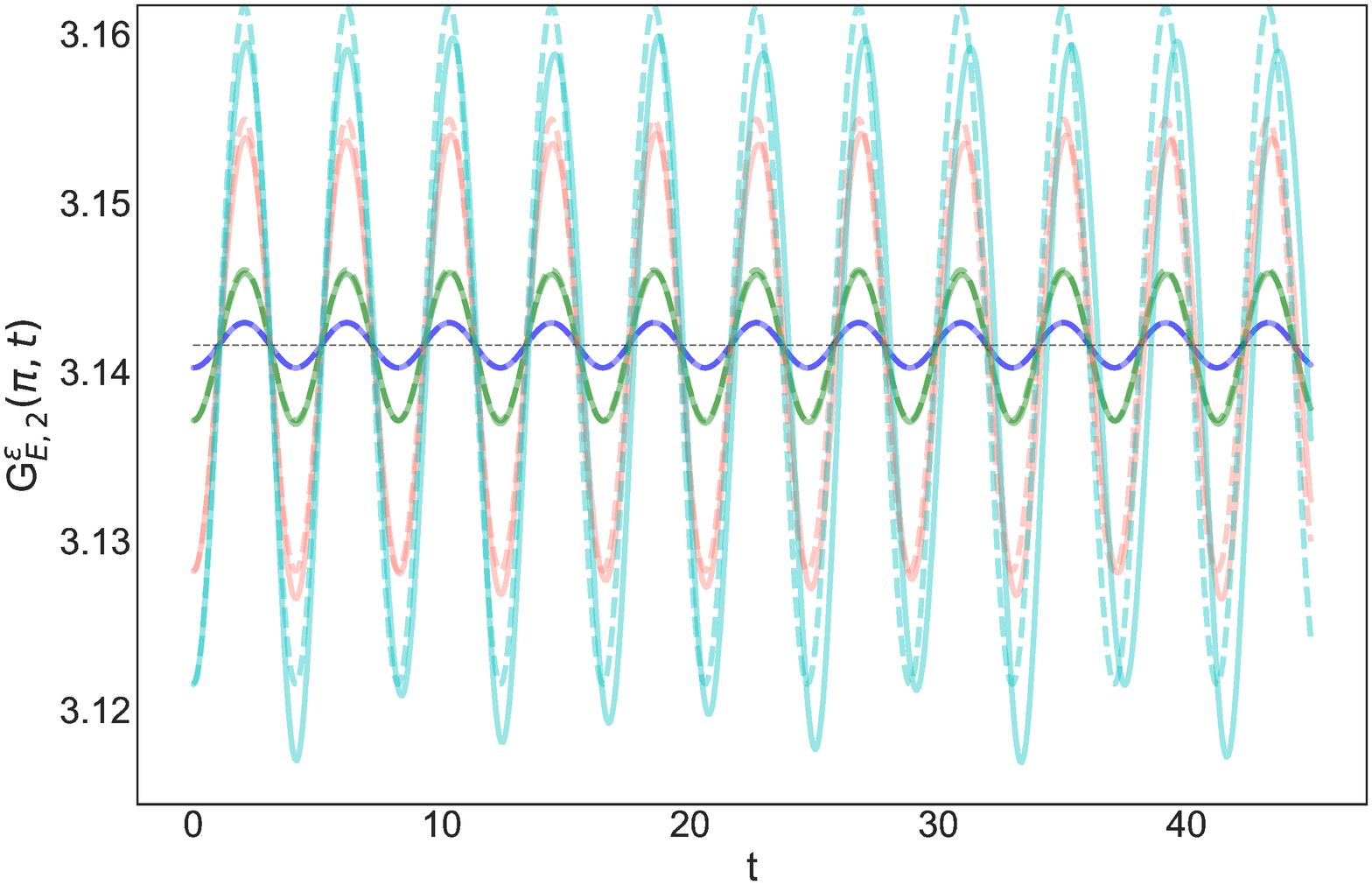}
  \textbf{a}   
     \includegraphics[scale=0.31]{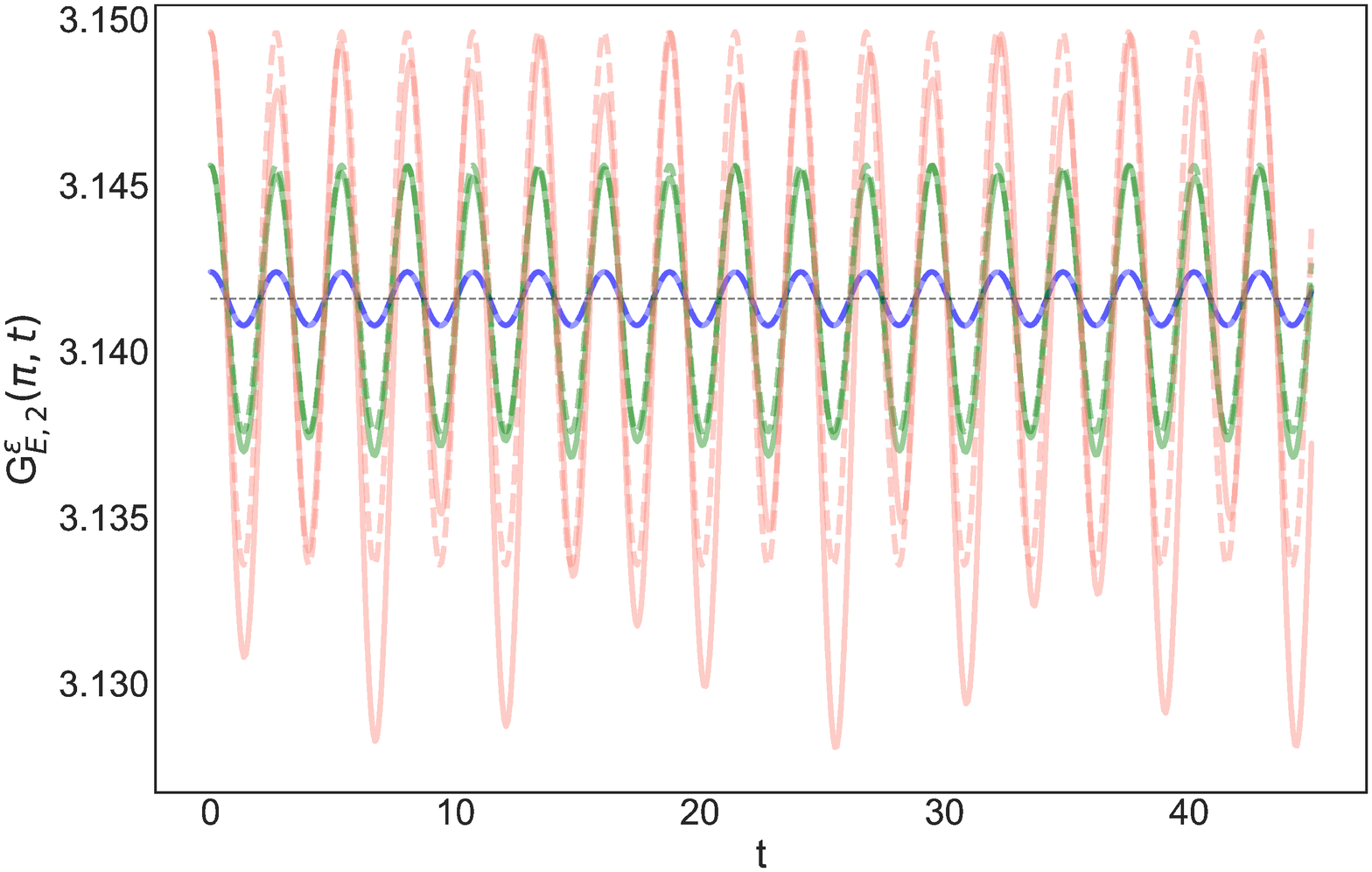}
 \textbf{b}    
      \includegraphics[scale=0.31]{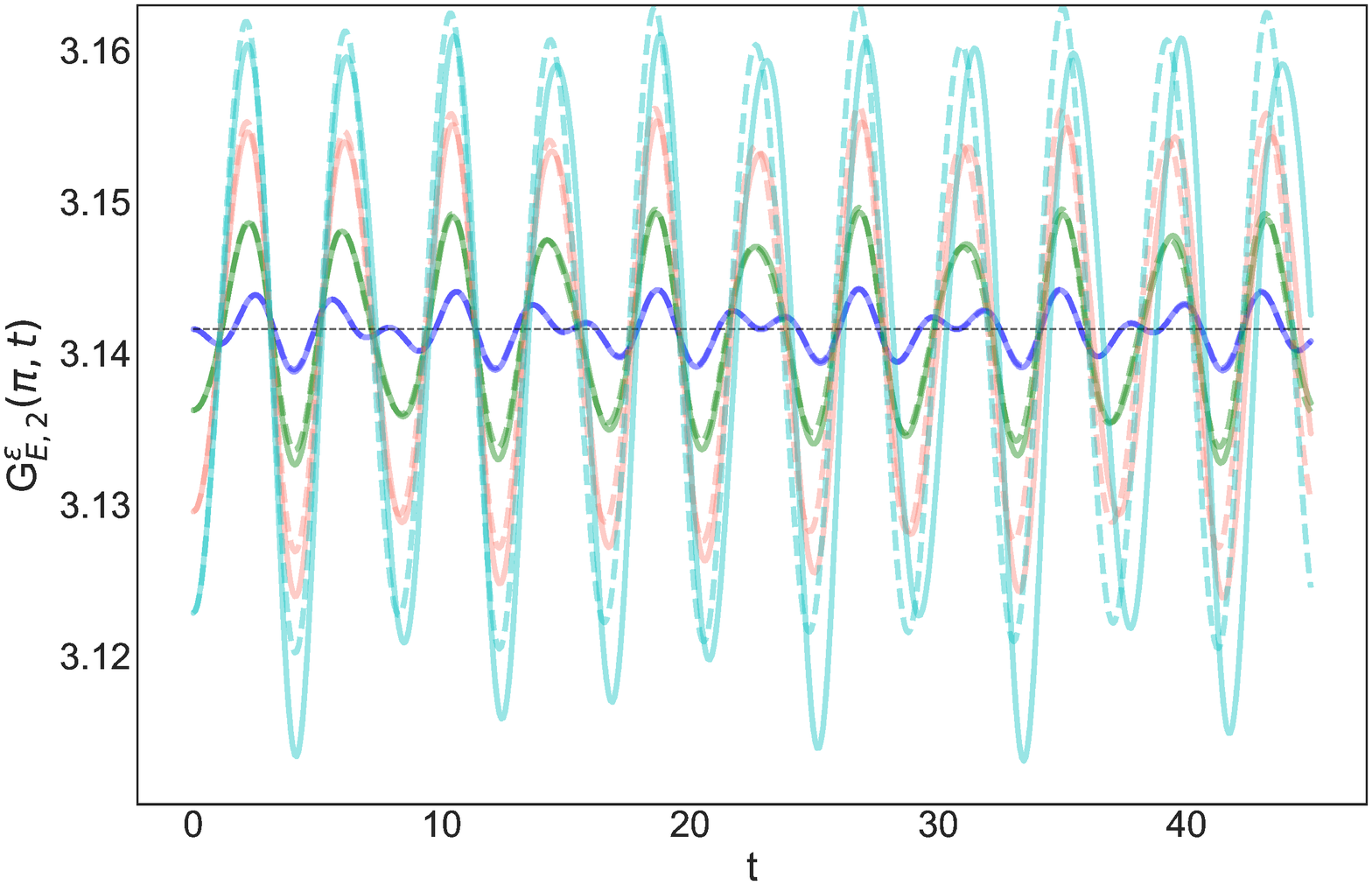}
 \textbf{c}      
       \caption{Run-up, i.e. second component of ${\cal G}^\epsilon_{E}(s,t)$ at $s = \pi$.
       Solid lines (common value of $\epsilon$; color, initial condition). Dashed lines correspond to linear evolution for same initial conditions.      
 (a) $\epsilon = 0.1$; blue: $(0.1,0,0,0)$, green: $(0.5,0,0,0)$,
 pink: $(1.0,0,0)$, light blue: $(1.5,0,0,0)$, see Fig. 2.
   (b) $\epsilon = 0.09$; blue: $(0,0.1, 0,0)$,  green: $(0,0.5,0,0)$, pink: $(0,1,0,0)$, see Fig. 3. 
   (c)  $\epsilon = 0.1$; blue: $(0.1,0.1,0,0)$, green: $(0.5,0.1,0,0)$, 
   pink: $(1,0.1,0,0)$, light blue: $(1.5,0.1,0,0)$, see Fig. 4.        }
    \label{Caso1Runup}
\end{figure}

These qualitative observations of the spatial patterns can be made more precise by examining the surface elevation 
at a given point of the domain as a function of time and the initial amplitude.  
We consider the second component of ${\cal G}^{\epsilon}_{E,T} (s,t)$ of \eqref{trunctated-GEtilde}, i.e. the height 
of the surface, 
at different points $s \in [0,\pi]$. The case of $s = \pi$ is of special interest as it gives us the run-up of the 
waves at the beach. 
In Figures \ref{Caso1Runup} (a), (b), (c)
we show the run-up for the different initial amplitudes considered
in the experiments of Figures   
\ref{fig:b2-b4-a2-a4-index2-1p5} (mode $n =2 $ near-monochromatic wave), 
\ref{fig:b2-b4-a2-a4-index4-1} (mode $n =4 $ near-monochromatic wave)
\ref{fig:b2-b4-a2-a4-index2and4-1p5}
(bichromatic wave)
respectively. 
In the case of the near-monochromatic waves, higher amplitudes lead to higher temporal variability
of the amplitude, and also to lower maxima and minima.
In the case of bichromatic waves we see higher temporal variability for smaller amplitudes.  
At $ s = 0 $ we see a similar behavior,  
although local maxima and minima 
of the height are closer for the amplitudes examined.

%
%


  
%
%



%
%




\section{Discussion}

We have derived a quadratic model for weakly nonlinear waves in a triangular channel with 
inclined walls of $45^o$ and presented calculations 
of phenomena such as the influence of weak nonlinearity in the wave run-up. 
The derivation may be valid in more general domains, assuming that a similar construction 
of harmonic functions related to the linear normal modes of the problem is possible in more general domains.  

Justifying the steps leading to the approximate model presented for the $45^0$ triangular domain 
and possible generalizations will require additional analysis, and we have pointed out some of the problems. 
The representation of the free surface we proposed appears to be new, but its generality remains to be seen.
Its use is justified heuristically by the fact that it leads to the linear theory (it uses 
an infinite set of parameters). At the same time it leads to  
a clear definition of the position of the free surface.     
In the $45^o$ domain the use of functions that are harmonic on the whole plane bypasses some of the 
regularity issues for Dirichlet-Neumann operators in domains with corners \cite{MW16}, and allows  
the definition of the free surface to go through. The existence of these functions is also behind a class of methods for the  
numerical evaluation of the Dirichlet-Neumann operator \cite{CS93,WV15}, and this is a question that could be examined further in the 
$45^o$ triangular domain. 
While the possibility of treating general domains is a significant motivation for this study, 
the question of nonlinear interactions in the geometries with known normal modes is also 
of interest as it also extends to three-dimensional domains. 

The quadratic model and its spectral truncations pose several interesting dynamical questions. 
Local and global well-posedness for the quadratic model is not clear at present.
Further numerical studies of spectral truncations should describe a wider range of bounded motions. 
The existence of nontrivial fixed points is a possible mechanism for unbounded growth in  
small spectral truncations. Such solutions seem tractable and can be also studied further. 
Extensions to cubic models 
may be possible. Clearly spectral models involve long intermediate calculations
of mode interaction coefficients, and this is a drawback of our approach. 
The exact expressions for the interaction coefficients presented suggest some heuristic simplifications
for short waves.

The model we presented does not have an apparent Hamiltonian structure.  
Work on a model with Hamiltonian structure is currently in progress. 
The Hamiltonian formalism involves approximation and 
symmetrization of a Dirichlet-Neumann operator for a perturbation of the flat surface domain. 
This operator was not discussed explicitly here, although we saw some operations seen in the computation 
of an approximate Dirichlet-Neumann operator, see Remark \ref{implicit-psi-evol}. The lowest order nonlinear
Hamiltonian model leads to an equation that is similar in structure to the one presented here but with 
more complicated coefficients. Thus while the Hamiltonian structure is a desirable feature, the present model 
appears to be simpler. The lack of Hamiltonian structure is a drawback but also leads to interesting dynamical questions. 
Also, from the point of view of applications, the general structure of the evolution equations 
is of importance as some details could be obtained using observations.

{\bf Acknowledgments.}

	The authors acknowledge partial support from grant PAPIIT IN100522. R. M Vargas-Maga\~na also acknowledges partial support from CONACyT-Postdoctoral fellowship call EPE2019(2) and the collective ``Cientific{\ae}s Mexican{\ae}s en el Extranjero.'' We also wish to thank Sebastien Fromenteau for valuable guidance with Python's visualization tools,
and Gerardo Ruiz Chavarr\'ia, Carlos Garc\'ia Azpeitia, and Noel F. Smyth for helpful discussions.

\section{Appendix}

We briefly discuss the dispersion relation and present some details of the computation of the mode interaction
coefficients.  

The dispersion relation can be analyzed using  
equations \eqref{sym-mode-alpha}, \eqref{antisym-mode-alpha} 
and the theory of Newton's method.  Details may be presented elsewhere,   
we here limit ourselves to showing the monotonicity of $\omega_n$.  

{\it Proof of Proposition \ref{Prop-dispersion}}:  
We first show monotonicity of $\omega_n$, $n \geq 2$.
Let 
$n $ even then by 
\eqref{disp-relation}
$\omega_{n+1}^2 > \omega_n^2$ is 
equivalent to $ \alpha_{n+1} \coth \alpha_{n+1}B > \alpha_n \tanh \alpha_n B $.
This holds by 
$ \alpha_{n+1}/{\alpha_n} > 1$, 
see  
\eqref{sym-mode-alpha},
\eqref{antisym-mode-alpha}, 
and $\tanh \alpha_m B < 1$, for all $m \geq 1$.  

We also need to show $\omega_{n+1}^2 > \omega_n^2$ for $n \geq 3$ and odd, 
equivalently 
\begin{equation}
\label{ratio-coth-comparison}
\frac{ \alpha_{2k +2}}{\alpha_{2k+1}} > 
\coth \alpha_{2k+2} B \coth \alpha_{2k+1} B. 
\end{equation}  

By \eqref{sym-cond}, 
\eqref{sym-mode-alpha}
we have 
\begin{equation}
\label{odd-intersect}
B \alpha_{2 k + 2} > k \pi  + \frac{3 \pi}{4},
\end{equation}
$\forall k \geq 0$, 
 by comparing solutions of equations \eqref{sym-cond} and $-\tanh \alpha B = 1$, 
 and using $\tanh \alpha B < 1$ and 
 increasing in $\alpha$, 
 and $- \tan \alpha B$ decreasing in $\alpha$. 
 Similarly, by  
\eqref{antisym-cond}, 
\eqref{antisym-mode-alpha} 
we have 
\begin{equation}
\label{even-intersect}
B \alpha_{2 k +1 } < k \pi  + \frac{\pi}{4},
\end{equation}
$\forall k \geq 1$, 
by comparing solutions of \eqref{antisym-cond} and $\tanh \alpha B = 1$, 
and using $\tanh \alpha B < 1$ and increasing, and $\tan \alpha B$ increasing. 
By \eqref{odd-intersect}, \eqref{even-intersect} we then have 
\begin{equation}
\label{ratio-comp}
\frac{ \alpha_{2k +2}}{\alpha_{2k + 1}} > \frac{4 k + 3}{4k+1} = 
1 + \frac{1}{2k + \frac{1}{2}}, \quad \forall k \geq 1.  
\end{equation}

On the other hand 
\begin{equation}
\label{coth-product}
\coth \alpha_{2k+2} B \coth \alpha_{2k + 1} B = 
\left( 1 + 
\frac{2 e^{-2B \alpha_{2k+1}}   } { 1 - e^{- 2 B \alpha_{2k+2}}  }  \right) 
\left( 1 + 
\frac{2 e^{-2B \alpha_{2k+1}}  }{ 1 - e^{- 2 B \alpha_{2k+1}} }  \right).
\end{equation} 
By 
\eqref{sym-mode-alpha},
\eqref{antisym-mode-alpha} we 
have 
\begin{equation}
\label{exp-est}
e^{-2B \alpha_{2k+1}}  < e^{- 2 k \pi}, \quad 
e^{-2B \alpha_{2k+2}}  < e^{- (2 k+1) \pi}, \quad \forall k \geq 1. 
\end{equation}
We simplify further by using $e^{- k } < (2 k)^{-1}$, $\forall k \geq 1$, and 
$3 < \pi$, then  \eqref{coth-product}, \eqref{exp-est}
lead to 
\begin{equation}
\label{coth-product-bound}
\coth \alpha_{2k+2} B \coth \alpha_{2k+1} B  <  1 + \frac{3}{ 8 k^6}, \quad \forall k \geq 1. 
\end{equation}
Comparing \eqref{ratio-comp}, \eqref{coth-product-bound} we obtain 
\eqref{ratio-coth-comparison}. 

It remains to show $\omega_2 > \omega_1$, equivalently
$ B \alpha_2 \tanh B \alpha_2 > 1$,  or 
\begin{equation}
\label{first-comparison}
B \alpha_2 > \frac{1}{1 - T_2}, \quad T_2 = \frac{2 e^{-2 B \alpha_2}}{1 + 2^{- 2 B \alpha_2}}. 
\end{equation}  
By \eqref{odd-intersect} we have 
$ B \alpha_2  > 3\pi/4$, therefore $T_2 < 2e^{-3 \pi/2}  < 2^{-3}$. 
Then $(1 - T_2)^{-1} < 8/7$, while $B \alpha_2 > 9/4$. 
\hfill $\square$

{\it Proof of Proposition 
\ref{approximate-volume-conserv}}:  
The integral $I$ left hand side of \eqref{gradient-vol-conserv} is 
$$ 
I = 
\frac{1}{2} \int_{-h}^{h}[( B + \epsilon \partial_y V(s,B)) 
( 1 + \epsilon \partial_s \partial_x V(s,B)) - ( s + \epsilon \partial_x V(s,B))
( \epsilon \partial_s  \partial_y V(s,B) )] \; ds. $$
We split into powers of $\epsilon$ 
\begin{equation}
\label{I-split}
I = I_0 + \epsilon I_1 + \epsilon^2 I_2,
\end{equation}
then 
$$ I_0 =  \frac{1}{2} \int_{-h}^{h} B \; ds = B^2. $$

The second term is 
\begin{eqnarray}
I_1 & = & \frac{1}{2} \int_{-h}^{h}[
-s \partial_s \partial_y V(s,B) + B \partial_s \partial_x V(s,B) + \partial_y V(s,B)
] \; ds \\
& =  & 
\int_{-h}^{h} \partial_y V(s,B) \; ds + \left[-s \partial_y V(s,B)\right]^{s=h}_{s=-h}
+ B\left[\partial_x V(s,B)\right]^{s=h}_{s=-h}.
\end{eqnarray}
Using the boundary condition for $V$ at $\partial {\cal W}$ we have  
$ [1,-1] \cdot \nabla V (h,B)  = 0$, and 
$[1,1] \cdot \nabla V (-h,B) = 0$ so that 
the boundary terms vanish.
Then 
\begin{equation}
I_1 = \int_{-h}^{h} \partial_y V(s,B) \; ds = 0,
\end{equation}
by Green's identity in the triangle defined by $[X,Y]$ and $\partial {\cal W}$, 
and $ \Delta V = 0$ in $\cal W$, ${\hat n} \cdot \nabla V = 0$ at $\partial {\cal W}$. 

The third integral in \eqref{I-split} is
\begin{eqnarray}
I_2 & = & \frac{1}{2} \int_{-h}^{h}[- (\partial_x V(s,B)) \partial_s \partial_y V(s,B) +
(\partial_y V(s,B)) \partial_s \partial_x V(s,B)] \; ds \\
& = & - \int_{-h}^{h} \partial_x V(s,B) \partial_s \partial_y V(s,B) \; ds + 
\left[ (\partial_x V(s,B)) (\partial_y V(s,B)) \right]^{s=h}_{s=-h}
\end{eqnarray}
and is not guaranteed to vanish.
\hfill $\square$


%
%
%
%
%
%

\end{document}